%% file: main.tex
\documentclass[sigconf]{acmart}
\AtBeginDocument{%
  }

\setcopyright{acmlicensed}

\copyrightyear{2025}
\acmYear{2025}
\setcopyright{cc}
\setcctype{by}
\acmConference[CCS '25]{Proceedings of the 2025 ACM SIGSAC Conference on
Computer and Communications Security}{October 13--17, 2025}{Taipei, Taiwan}
\acmBooktitle{Proceedings of the 2025 ACM SIGSAC Conference on Computer
and Communications Security (CCS '25), October 13--17, 2025, Taipei,
Taiwan}\acmDOI{10.1145/3719027.3765067}
\acmISBN{979-8-4007-1525-9/2025/10}



\input{packages}

\input{notions}

\begin{document}

\title{Systematic Assessment of Tabular Data Synthesis}


\author{Yuntao Du}
\affiliation{%
    \institution{Purdue University}
    \city{West Lafayette}
    \country{USA}
}
\email{ytdu@purdue.edu}

\author{Ninghui Li}
\affiliation{%
    \institution{Purdue University}
    \city{West Lafayette}
    \country{USA}
}
\email{ninghui@purdue.edu}


\begin{abstract}
Data synthesis has been advocated as an important approach for utilizing data while protecting data privacy. In recent years, a plethora of tabular data synthesis algorithms (\ie synthesizers) have been proposed. 
Some synthesizers satisfy Differential Privacy, while others aim to provide privacy in a heuristic fashion. 
A comprehensive understanding of the strengths and weaknesses of these synthesizers remains elusive due to drawbacks in evaluation metrics and missing head-to-head comparisons of newly developed synthesizers that take advantage of diffusion models and large language models with state-of-the-art statistical synthesizers.

In this paper, we present a systematic evaluation framework for assessing tabular data synthesis algorithms. 
Specifically, we examine and critique existing evaluation metrics, and introduce a set of new metrics in terms of fidelity, privacy, and utility to address their limitations. 
We conducted extensive evaluations of 8 different types
of synthesizers on 12 real-world datasets and identified some interesting findings, which offer new directions for privacy-preserving data synthesis.

\end{abstract}

\maketitle

\input{1_intro}

\input{2_framework}
\input{3_fidelity}

\input{4_privacy}

\input{5_utility}

\input{6_synmeter}

\input{7_exp}
\input{0_related}
\input{8_conclusion}

\begin{acks}
We thank the anonymous reviewers for their constructive comments.  This work was funded in part by the National Science Foundation (NSF) Awards CNS-2247794 and IIS-2229876.  Any opinions, findings, and conclusions or recommendations expressed in this material are those of the authors and do not necessarily reflect the views of the sponsors.
\end{acks}

\bibliographystyle{ACM-Reference-Format}
\balance
\bibliography{ref}

\input{9_appendix}

\end{document}

%% file: packages.tex
\usepackage{multirow}
\usepackage{booktabs}
\usepackage{subfigure}
\usepackage{array}
\usepackage{pifont}

\usepackage{graphicx}
\usepackage{bbm}
\usepackage{nicefrac}       
\usepackage{amsmath} 
\usepackage{enumitem}
\usepackage{caption}        
\usepackage{mathtools}      
\usepackage{amsfonts}
\usepackage{mathtools}
\usepackage{pifont}
\usepackage{xcolor}
\usepackage{xurl}
\usepackage{balance}
\usepackage[flushleft]{threeparttable}
\definecolor{mycitecolor}{RGB}{0,0,153}
\definecolor{mylinkcolor}{RGB}{179,0,0}
\definecolor{myurlcolor}{RGB}{255,0,0}
\definecolor{darkgreen}{RGB}{0,170,0}
\definecolor{darkred}{RGB}{200,0,0}

\usepackage{hyperref}
\usepackage{cleveref}

\setlist[itemize]{leftmargin=*,noitemsep, topsep=0pt}

\usepackage{xspace}
\usepackage{multirow}
\usepackage{amsthm}
\theoremstyle{plain}
\newtheorem{definition}{Definition}

\usepackage{dsfont}

\usepackage[toc,page,header]{appendix}
\usepackage{minitoc}

%% file: notions.tex
\newcommand{\ie}{\emph{i.e., }}
\newcommand{\eg}{\emph{e.g., }}

\newcommand{\mypara}[1]{\smallskip\noindent\textbf{#1.} \xspace}

\newcommand{\myquestion}[1]{\smallskip\noindent\textbf{#1?} \xspace}

\newcommand{\mymethod}{\ensuremath{\mathsf{SynMeter}}\xspace}

\newcommand{\half}{\texttt{HALF}\xspace}
\newcommand{\hist}{\texttt{HISTOGRAM}\xspace}
\newcommand{\self}{\texttt{SELF}\xspace}

\newcommand{\revision}[1]{\textcolor{black}{#1}}

\newcommand{\halfcheckmark}{
  \ding{51}
  \kern-0.7em\raisebox{1.2ex}{\rotatebox{-45}{\rule{0.4em}{0.08em}}}
}



%% file: 1_intro.tex
\section{Introduction}

Data-driven decision-making has emerged as the prevailing approach to advance science, industrial applications, and governance, creating the necessity to share and publish tabular data.  
At the same time, growing concerns about the privacy breaches caused by data disclosure call for data publishing approaches that preserve privacy. 
One increasingly advocated and adopted approach to mitigate privacy risks while sharing data is to release synthetic data. 
Ideally, synthetic data can effectively fit any data processing workflow designed for the original data without privacy concerns. 
Data synthesis initiatives have been promoted by academia~\citep{arxiv21benmarkdpsyn}, organizations~\citep{oecd23}, and government~\citep{uscensus18}.

In this paper, we study data synthesis algorithms for tabular data, which we call \textbf{synthesizers}.  
In recent years, a plethora of synthesizers have been proposed, which can be roughly categorized into two groups: statistical and deep generative. Statistical synthesizers use low-order marginals to create synthetic datasets that match real distributions. They were the best-performing algorithms in NIST competitions~\citep{nist18,nist20}.  
Deep generative synthesizers, on the other hand, learn the data distribution from real data and generate synthetic instances by sampling from the learned distribution. With the recent development in deep generative models (\eg diffusion models~\citep{nips20ddpm} and large language models (LLMs)~\citep{nips17attention,openai2019llm}), new synthesizers are proposed to extend these successes to the realm of tabular data synthesis.  

While recent studies have achieved compelling results in synthesizing tabular data, a comprehensive understanding of the strengths and weaknesses of different synthesizers remains elusive.  
We are unaware of any head-to-head comparison of the best statistical synthesizers~\cite{icml19pgm,mckenna21mst,usenix21privsyn} against the best deep generative ones~\cite{icml23tabddpm,iclr23great,iclr23stasy}.  
In addition, there is a lack of principled and widely accepted evaluation metrics for data synthesis.  
It is known that evaluating synthesizers is inherently difficult~\cite{iclr16evalgen,ai22metric}, and qualitative evaluation of tabular data through visual inspection is also infeasible.

Since synthetic data is proposed to protect the privacy of real data, privacy should be an indispensable part of the evaluation. While many \textit{differentially private (DP)} synthesizers~\cite{tods17privbayes,mckenna21mst,usenix21privsyn,vldb22aim} have been proposed with provable privacy guarantees, they often suffer from significant utility loss~\cite{arxiv21benmarkdpsyn,icml22robinhood,sp24ppds}. 
Therefore, a parallel line of research focuses on what we call \textit{heuristically private (HP)} synthesizers~\cite{vldb18tablegan,acml21ctgabgan,icml23tabddpm,iclr24tabsyn,iclr23great,iclr23stasy}, which do not integrate DP for model learning.
These methods typically rely on privacy evaluation metrics that measure the similarity between the synthetic dataset and the input dataset to empirically evaluate privacy risks. 
These metrics are widely used both in academia and industry as default privacy evaluation protocols (see~\cite{arxiv23privcymetric} for detailed analysis).
We note that these metrics are \textit{syntactic} in the sense that they are properties of a pair of input/output datasets and are independent of the synthesis algorithm that generates the output dataset.  
As a result, they suffer from several weaknesses in privacy evaluation metrics.  
We conducted experiments to analyze their effectiveness and found that these metrics are highly unstable and cannot distinguish different levels of privacy leakage.

The above concerns motivate us to design a systematic evaluation framework for data synthesis to elucidate the current advancements in this field.
Specifically, we examine, characterize, and critique the commonly used metrics and propose a set of new evaluation metrics. Our assessments unfold along three main axes:

\begin{itemize}
    \item \textit{Fidelity.} 
    To address the heterogeneity of tabular data, we present a fidelity metric based on Wasserstein distance. This metric offers a unified way to evaluate numerical, discrete, and mixed data distributions under the same criteria.
    \item \textit{Privacy.} 
    We identify the inadequacy of existing syntactic privacy evaluation metrics and the ineffectiveness of membership inference attacks by conducting comparison studies. We also propose a new privacy evaluation metric called membership disclosure score 
    to gauge the empirical privacy risks of synthesizers. 
    \item \textit{Utility.} 
    We advocate two tasks for assessing the utility of synthesizers: machine learning prediction and range (point) query. To eliminate the inconsistent performance caused by the choice of different machine learning models, 
    we present a utility evaluation metric that quantifies the distributional shift between real and synthetic data.
\end{itemize}

\mypara{\mymethod} We implement a systematic evaluation framework called \mymethod to support the assessment of data synthesis algorithms with the proposed evaluation metrics. 
\mymethod incorporates the model tuning phase, which eases hyperparameter selection and consistently improves the performance of synthesizers for fair comparison.  
With a modular design, \mymethod can easily integrate additional synthesizers and datasets by implementing new functional codes in the relevant modules.  
Our code is publicly available\footnote{\href{https://github.com/zealscott/SynMeter}{https://github.com/zealscott/SynMeter}}, facilitating researchers to tune, assess, or benchmark new tabular data synthesis algorithms.

\mypara{Main Findings} 
Extensive experiments have been conducted on a wide range of state-of-the-art heuristically private (HP) and differentially private (DP) synthesizers (listed in Section~\ref{sec:exp_setup}) over 12 real-world datasets.
Our evaluation finds that no single synthesis algorithm emerges as the definitive leader. 
Synthesizers struggle to balance competitive performance with robust privacy measures, exhibiting a pronounced performance disparity between HP and DP synthesis algorithms. 
Specifically, we have the following findings:
\begin{itemize}
    \item \textit{Diffusion models are surprisingly good at synthesizing (non-private) tabular data.} 
    Although they were originally introduced for image generation, our evaluation indicates their impressive capability of synthesizing highly authentic tabular data. 
    For instance, the HP diffusion-based synthesizer (\ie TabDDPM~\cite{icml23tabddpm}) surpasses its counterparts and nearly reaches the empirical upper bound in terms of both fidelity and utility. 
    However, it suffers from significant membership privacy risks, and directly applying differential privacy to it would negate its advantages, highlighting a critical challenge for further exploration. 
    \item \textit{Statistical methods are still competitive synthesizers.} 
    State-of-the-art statistical methods consistently outperform all deep generative synthesizers when DP is required, especially when the privacy budget is small (\eg $\epsilon=0.5$). Impressively, even in heuristically private settings (\ie $\epsilon=\infty$), these statistical approaches still maintain competitive performance and empirical privacy protection over many sophisticated synthesizers like TVAE~\cite{nips19ctgan}.
    \item \textit{LLMs are semantic-aware synthesizers.} 
    LLM-based method (\eg REaLTabFormer~\cite{arxiv23realtabformer}) excels at generating realistic tabular data when the input dataset consists of rich semantic attributes, establishing a new paradigm in tabular data synthesis. 
    \item \textit{CTGAN is not an effective tabular data synthesizer.}
    Despite CTGAN~\cite{nips19ctgan} being widely recognized as a strong baseline in data synthesis, a closer examination using proposed metrics reveals that it struggles to learn marginal distributions, which leads to unsatisfactory results on complex tabular data.
\end{itemize}

\mypara{Related Studies}
Several studies~\cite{arxiv21benmarkdpsyn,sp24ppds,arnold2020really,bowen2019comparative} have benchmarked tabular synthesis algorithms. However, they all focus on DP synthesizers, neglecting extensively studied HP synthesizers. 
In particular, they did not include the recently proposed deep generative synthesizers using diffusion models or LLM~\cite{icml23tabddpm,iclr23great,iclr23stasy}, which we found to be very promising in our studies. 
Furthermore, they directly use existing metrics for comparisons, while we identify the limitations of these metrics and propose a new set of metrics as well as a systematic assessment for tabular data synthesis.
We refer to Section~\ref{sec:related} for a detailed discussion of related work.

%% file: 2_framework.tex
\section{A Framework for Data Synthesis Evaluation}
\label{sec:framework}

\subsection{Evaluation Criteria}

Given a dataset $D$ sampled from an underlying distribution $\mathbb{D}$, $\mathsf{A} \gets \mathcal{T}(D)$ denotes that the synthesizer $\mathsf{A}$ is learned by running the training algorithm $\mathcal{T}$ on $D$. The synthesizer $\mathsf{A}$ generates a synthetic dataset $S$ to replace $D$ for publishing. We consider three classes of properties for synthesizers:
\begin{itemize}
    \item \textbf{Fidelity}. As a substitute for real data, the distribution of the synthetic dataset should be close to $\mathbb{D}$.  Since $\mathbb{D}$ is often unknown, fidelity is measured by the similarity between the input data $D$ and the synthetic data $S$.  
    This can be assessed by comparing $S$ to either the training set $D_{\text{train}}$ or the test set $D_{\text{test}}$ of $D$.
    \item \textbf{Privacy}. 
    Using synthetic data is usually motivated by the desire to protect the input dataset. Some training algorithms $\mathcal{T}$ are designed to satisfy Differential Privacy (DP)~\citep{dwork_dp}, we refer to these as DP synthesizers. 
    However, satisfying DP under reasonable parameters may result in poor performance. Some synthesizers do not satisfy DP and aim to protect privacy empirically. We call these Heuristically Private (HP) synthesizers. As a result, privacy evaluation metrics are essential for evaluating the privacy risks of HP synthesizers.
    \item \textbf{Utility}. Synthetic data is often used to replace real datasets for downstream tasks. High fidelity may not necessarily be needed if it achieves good utility for these tasks. Hence, utility is useful to measure the effectiveness of synthesizers for common tasks. 
\end{itemize}

\subsection{Evaluation Pipelines}  
\label{sec:pipeline}
The \mymethod pipeline consists of four phases: data preparation, model tuning, model training, and model evaluation. 

The \textit{data preparation} phase preprocesses data for learning algorithms. Here, we assume no missing values in the original data, since the missing values problem~\citep{pigott01review} is orthogonal to data synthesis. In this phase, statistical methods select low-dimensional marginals to serve as compact representations for capturing data distributions. Deep generative models apply standard data processing techniques like data encoding and normalization. 

The goal of \textit{model tuning} phase is to select the optimal hyperparameters for data synthesizers. We use the proposed tuning objective in Section~\ref{sec:implements} for hyperparameter selections.

The \textit{model training} phase focuses on model learning with tuned hyperparameters. Various generative models implement different architectures and optimization objectives. 

In the \textit{model evaluation} phase, the trained model generates synthetic data, which is used to evaluate fidelity, privacy, and utility.


%% file: 3_fidelity.tex
\section{Evaluation Metrics for Data Synthesis}

\subsection{Fidelity Evaluation}\label{sec:fidelity}
In this section, we first review existing fidelity measurements, identify their limitations, and introduce a fidelity metric that addresses these limitations.

\mypara{Existing Metrics and Limitations}
Existing fidelity metrics can be categorized into three groups: low-order statistics~\citep{icml19pgm}, likelihood fitness~\citep{nips19ctgan}, and evaluator-dependent metrics~\citep{snoke2018general}. 
The main issue with low-order statistics is the lack of versatility. Each type of marginal distribution requires a specific statistical measure, complicating comprehensive comparisons across different attribute types. 
Likelihood fitness assesses how well synthetic data aligns with a known prior distribution. Although this is a natural approach for assessing fidelity, it becomes problematic when the prior distribution is unknown or complex, as is often the case in real-world datasets.
Evaluator-dependent metrics, on the other hand, rely heavily on auxiliary evaluators, which require careful calibration to ensure meaningful comparisons across diverse datasets and synthesizers. A more detailed discussion of existing fidelity evaluation metrics can be found in Appendix~\ref{appendix:existing_fidelity_metrics}.

\mypara{Proposed Metric: Wasserstein Distance}
We opt for Wasserstein distance to measure the distribution discrepancies between synthetic and real data.
Originating from optimal transport theory~\citep{peyre2019opt}, the Wasserstein distance provides a structure-aware measure of the minimal amount of work required to transform one distribution into another. 
Formally, let $\mathbf{P}=(p_1, p_2, \ldots p_n)$ and $\mathbf{Q}=(q_1, q_2, \ldots q_n)$ be the two probability distributions, and $\mathbf{C}$ be a matrix of size $n\times n$ in which $\mathbf{C}_{ij} \geq 0$ is the cost of moving element $i$ of $\mathbf{P}$ to element $j$ of $\mathbf{Q}$ ($\mathbf{C}_{ii} = 0$ for all elements). The optimal transport plan $\mathbf{A}$ is:
{\setlength{\abovedisplayskip}{3pt}
\setlength{\belowdisplayskip}{3pt}
\begin{equation} \label{equ:wassertein_opt}
\begin{split}
& \min_{\mathbf{A}} \quad \langle \mathbf{C}, \mathbf{A}\rangle \\
\text { s.t. } & \mathbf{A} \mathds{1} = \mathbf{P}, \quad  \mathbf{A}^\top \mathds{1} = \mathbf{Q},
\end{split}
\end{equation}}where $\langle\cdot,\cdot\rangle$ is inner product between two matrices, $\mathds{1}$ is a vector of all ones. Let $\mathbf{A}^*$ be the solution to the above optimization problem. Wasserstein distance is defined as:
{\setlength{\abovedisplayskip}{3pt}
\setlength{\belowdisplayskip}{3pt}
\begin{equation} \label{equ:wassertein}
\mathcal{W}(\mathbf{P},\mathbf{Q}) = \langle \mathbf{C}, \mathbf{A}^*\rangle.
\end{equation}}Now we can use the Wasserstein distance to define the fidelity.
\begin{definition}[Wasserstein-based Fidelity Metric]
    Let $v$ be a set of marginal variables, and $V=\{v\}$ is the collection of marginal variable sets. $f(v,D)$ is the marginal extraction function that derives the corresponding marginal distribution of $v$ from distribution $D$. Let $D$ and $S$ be the empirical distribution of real and synthetic data, respectively. The fidelity of synthesis algorithm $\mathsf{A}$ is:
    {\setlength{\abovedisplayskip}{3pt}
    \setlength{\belowdisplayskip}{3pt}
    \begin{equation}
        \mathrm{Fidelity} (\mathsf{A}) \triangleq \frac{1}{|V|} \sum_{v\in V} \mathcal{W}(f(v, D), f(v,S)).
    \end{equation}}
\end{definition}
The smaller Wasserstein distance indicates higher fidelity.

\mypara{Determining Cost Matrix}
The Wasserstein distance requires the predefined cost matrix $\mathbf{C}$, which encapsulates the ``cost'' of transitioning from one element to another. 
For $k$-way marginal distributions $\mathbf{P}$ and $\mathbf{Q}$, the cost matrix is formulated by summing the pairwise distances between corresponding elements:
{\setlength{\abovedisplayskip}{3pt}
\setlength{\belowdisplayskip}{3pt}
\begin{equation}
    \mathbf{C}_{ij} = \sum_{r=1}^k d (v_i^r, v_j^r).
\end{equation}}Here, $v_i, v_j \in \ensuremath{\mathbb{R}}^k$ are the element located in $i$ and $j$ in $k$-way probability distributions. The distance $d(\cdot, \cdot)$ is tailored to the nature of the attributes, differing for numerical and categorical values:
{\setlength{\abovedisplayskip}{3pt}
\setlength{\belowdisplayskip}{3pt}
\begin{equation}
\label{equ:cost_function}
    d (v_i^r, v_j^r) = 
        \begin{cases}
            ||v_i^r -v_j^r||_1, & \text{if numerical}\\
            
             \infty \ (\text{if } v_i^r \neq v_j^r), 1 \ (\text{if } v_i^r=v_j^r).  &  \text{if categorical} 
        \end{cases}
\end{equation}}

\mypara{Wasserstein Distance for Categorical Attributes}
Wasserstein distance is typically defined for metric spaces and is well-suited for numerical attributes. 
We note that the above definition for categorical attributes is equivalent to the computation of total variation distance~\citep{icml23tabddpm} and contingency similarity~\citep{SDV} for one/two-way marginal distributions.
Additionally, it is also feasible to assign semantic distance for categorical attributes~\citep{icnss21mgd}, we omit it because it depends on the specific context, and most synthesizers do not model the semantics in tabular data.


\mypara{Merits of Wasserstein-based Fidelity Metric} Wasserstein distance offers several advantages for evaluations:
(i) Faithfulness. It is a natural and structure-aware statistic measure for analyzing distribution discrepancies and generalizing existing metrics like total variation distance.
(ii) Universality. It accommodates both numerical and categorical attributes and extends to any multivariate marginals under the same criterion, facilitating the evaluation of heterogeneous types of marginals.


%% file: 4_privacy.tex
\subsection{Privacy Evaluation}\label{sec:privacy}
In this section, we examine the popular privacy evaluation methodologies for synthesizers and propose a novel privacy metric: membership disclosure score.  

\mypara{Existing Metrics and Limitations}
A popular approach to assess privacy risk for HP synthesizers is to compare the similarity between input and synthetic data, with higher similarity suggesting greater information leakage.
We call these metrics \textit{syntactic} because they  
consider only the input dataset and synthetic dataset, not the algorithm used to generate synthetic data. 
The most popular syntactic metric is Distance to Closest Records (DCR)~\citep{acml21ctgabgan}, which looks
at the distribution of the distances from each synthetic data point to its nearest real one and uses the 5th percentile of this distribution as the privacy score. DCR and other similar metrics are widely used in academia~\citep{Walia2020SynthesisingTD} and industry~\citep{aws22,gretel23}, and have become the conventional evaluation metric for HP synthesizers~\citep{arxiv23privcymetric}. 

We point out that syntactic privacy evaluation notions that are independent of the underlying algorithm are fundamentally flawed. For example, a synthesis algorithm that applies the same fixed perturbation to every record could produce a synthetic dataset that is quite different from the input dataset, resulting in a good privacy score under a syntactic metric, even though the input dataset could be easily reconstructed from the synthetic dataset.

Membership inference attacks (MIAs) have been widely used for privacy evaluation in machine learning (especially classification models)~\citep{sp17membership}. 
A few MIAs against tabular synthesizers have been proposed: Groundhog~\citep{usenix22tab_mia}, TAPAS~\citep{houssiau2022tapas}, and MODIAS~\citep{aistats23}. 
Our comparison studies in Section~\ref{sec:exp_effective_metrics} demonstrate that these MIA algorithms are limited in effectiveness: they fail to distinguish different levels of privacy leakage in some situations.
The detailed analysis of existing privacy evaluation metrics is in Appendix~\ref{appendix:existing_privacy_metrics}.

\mypara{Proposed Metric: Membership Disclosure Score (MDS)}
We propose a new evaluation metric to assess the membership disclosure risks of synthesizers, which is inspired by both DCR and MIAs. 
The intuition behind MDS is that the inclusion or exclusion of each record $x \in D$ during training may lead to different behaviors of the synthesizer $\mathsf{A}$, which can be measured as a function of $x, D,$ and $\mathsf{A}$.  We use the maximum value for any $x$ as the measure of privacy leakage of applying $\mathsf{A}$ to $D$. 
Specifically, we first define the disclosure risk of one record as follows.

\begin{definition}[Disclosure Risk of One Record]
    Let $O_D$ be the synthesizer $\mathsf{A}$'s output distribution when trained with dataset $D$, $\mathcal{M}$ is a distribution distance measurement, which is non-negative and symmetric. 
    The disclosure risk of record $x\in D$ is given by:
    {\setlength{\abovedisplayskip}{3pt}
    \setlength{\belowdisplayskip}{3pt}
    \begin{equation}\label{equ:disclosure}
    \mathrm{DS}(x, \mathsf{A}, D)\triangleq {\mathbb{E}}_{H \subset D \backslash x} \big [\mathcal{M} (O_H \| O_{H\cup \{x\}})\big ],
    \end{equation}}where $H$ is the subset of training instances that are i.i.d sampled from $D \backslash x$. The expectation is taken for the i.i.d sampling of $H$ and the randomness in synthesizer $\mathsf{A}$.
\end{definition}

Our privacy definition compares the difference between two expected output distributions for a given record $x$.  Unfortunately, the above computation is intractable: even the synthesizer's output distribution is not analytically known. To simplify the situation, we instead instantiate $\mathcal{M}$ to measure the closeness between $x$ and the empirical distribution of the synthetic data:
{\setlength{\abovedisplayskip}{3pt}
\setlength{\belowdisplayskip}{3pt}
\begin{equation*}
    \widehat{\mathrm{DS}}(x, \mathsf{A}, D) \triangleq \underset{\substack{H \subset D \backslash x \\ S\sim O_H \\ S^\prime\sim O_{H\cup \{x\}}}}
    {\mathbb{E}} \big [|\mathrm{dist}(x, S) - \mathrm{dist}(x, S^\prime)|   \big ].    
\end{equation*}}Here, $S$ is the synthetic dataset generated from $O_H$,
$\mathrm{dist}(x, S)$ denotes the nearest distance (under $l_1$ norm) between record $x$ and synthetic dataset $S$. (Empirically, we find that the difference between using $l_1$ and $l_2$ distance is negligible.) 
However, directly computing the above equation is computationally expensive because it requires training models on paired subsets $H$ and $H\cup \{x\}$ for every record $x$.
To address this, we employ shadow training techniques~\cite{sp17membership,sp22lira} commonly used in MIAs.
Specifically, we train $m$ synthesizers (\ie shadow models) on independently sampled subsets $H_1,...,H_m$ of equal size $|H_i| = \lfloor \frac{1}{2} |D|\rfloor$. 
To calculate the disclosure risk of $x$, we divide these shadow models into two groups: one trained on subsets where $x\in H$, and the other where $x\notin H$. 
For each model trained on these subsets, we randomly generate $n$ synthetic datasets and take the average nearest distance to $x$. 
Finally, we define the privacy risk of a synthesizer $\mathsf{A}$ on $D$ to be the \textit{maximum} disclosure risk across all training data:
\begin{definition}[Membership Disclosure Score]
Let $S$ be the sampled synthetic data from the synthesizer's output distribution $O_H$. The membership disclosure score of $\mathsf{A}$ is:
{\setlength{\abovedisplayskip}{3pt}
\setlength{\belowdisplayskip}{3pt}
\begin{equation}
\begin{split}
\label{equ:approx_ds}
    \mathrm{MDS}(\mathsf{A}) \triangleq 
    \max_{x\in D} \big | 
    \underbrace{{\mathbb{E}}_{H\subset D, S\sim O_{H\cup \{x\}}} [\mathrm{dist}(x, S)]}_{\text {closeness of } x \text { when trained with } x}
    - \\
    \underbrace{{\mathbb{E}}_{H \subset D \backslash x, S^\prime\sim O_{H}} [\mathrm{dist}(x, S^\prime)]}_{{\text {closeness of } x \text { when \textbf{not} trained with } x}}
    \big|  ,
\end{split}
\end{equation}}
\end{definition}

We train $m=20$ shadow models and generate $n=100$ synthetic datasets per model to compute MDS for all synthesizers.
We analyze the effectiveness, stability, and efficiency of MDS in Section~\ref{sec:exp_effective_metrics}.

We emphasize that MDS is not syntactic, as it depends on an input dataset $D$ and the synthesis algorithm $\mathsf{A}$, but it is independent of any synthetic dataset generated by a single run of the synthesizer. 

\mypara{Discussion}
We choose distance measurements to compute MDS for the following reasons:
(i) For many generative models, the distance between a target record $x$ and its closest synthetic data record is related to the probability density of $x$ (which is related to the loss of the model on $x$). Usually, the density is smooth. Thus, when $x$ has higher density, it is more likely that data records closer to $x$ are generated. Using distance instead of density (or loss) in MDS has the advantage that it does not require a way to compute the density of a given data record, which is difficult for some synthesizers (\eg GANs~\cite{nips19ctgan}). 
Instead, MDS requires only the synthesizers to output synthetic datasets, which all synthesizers must do.
(ii) The distance between real and synthetic data points is used in DCR~\cite{acml21ctgabgan}, a widely adopted privacy evaluation metric for tabular data synthesis. DCR has been employed in many (if not all) state-of-the-art HP synthesizers~\cite{iclr24tabsyn,icml23tabddpm,iclr23great}. 
We think one reason that DCR and other syntactic privacy metrics are so popular is their intuitive nature: they quantify the similarity between synthetic datasets and real datasets as privacy risks. 
Therefore, when we design MDS, we aim to come up with something conceptually similar (namely using distance to the closest data point), yet more aligned with the spirit of MIAs to avoid the pitfalls of DCR.
(iii) Empirical results in Section~\ref{sec:exp_effective_metrics} on both HP and DP synthesizers demonstrate that MDS outperforms both DCR and existing MIAs in effectively capturing privacy risks. 
We thus believe that MDS represents an advancement in the state of the art for privacy evaluation metrics in tabular data synthesis.

\revision{We note that MDS focuses on membership privacy; however, membership privacy implies bounds on other privacy leakages. 
For example, bounding membership leakage limits attribute inference of tabular data~\citep{ccs22attribute} because it bounds the ability to tell the difference when the sensitive attribute values of one record are changed. 
We also note that differential privacy (DP) is equivalent to no membership leakage under the assumption that memberships of different records are mutually independent~\citep{ccs13membership}. 
As DP is generally acceptable to be sufficient, we argue that addressing only membership leakage, as MDS does, is acceptable. }


\mypara{Limitations of MDS}
Although we find MDS to be effective in assessing the privacy risks of the synthesizers studied, we note that it has its own limitations. 
For instance, MDS can be tricked by carefully designed pathological synthesizers and should not be used as the only measure where privacy is paramount.
\revision{In addition, MDS requires training multiple shadow models to estimate the disclosure risks. 
This can pose a challenge when assessing large synthesizers like GReaT, which involve fine-tuning entire LLMs.
We provide a more detailed discussion of its limitations in Appendix~\ref{appendix:diss_mds}.}

%% file: 5_utility.tex
\subsection{Utility Evaluation}\label{sec:utility}

\mypara{Existing Metrics and Limitations}
Machine learning efficacy~\citep{nips19ctgan} has emerged as the predominant utility metric for data synthesis. It first chooses a machine learning model (\ie evaluator), then assesses the testing accuracy on real data after training the evaluator on synthetic datasets. 
However, there is no consensus on which evaluator should be used for evaluation. 
Different evaluators yield varying performance outcomes on synthetic data, and no single model consistently achieves the best performance across all datasets. (We show the case in Appendix~\ref{appendix:existing_utility_metrics}.)

\mypara{Advocated Metrics: Machine Learning Affinity (MLA) and Query Error}
\revision{To accurately reflect the performance degradation caused by the distribution shift of synthetic data~\citep{iclr21affinity}, prior work~\cite{royal2022syn_survey} suggests using various machine learning models for evaluation but does not provide a detailed approach. 
We extend this idea by measuring the relative performance gap as the utility metric:}

\begin{definition} [Machine Learning Affinity]
    Let $\mathcal{E}$ be a set of candidate machine learning models (i.e., evaluators), let $e_{D_\text{train}}$ and $e_{S}$ be evaluators trained on real training data $D_\text{train}$ and synthetic data $S$, $acc(e,D_{\text{test}})$ denotes the evaluator’s accuracy (F1 score or RMSE) when performed on test dataset $D_{\text{test}}$. The MLA of synthesizer $\mathsf{A}$ is given by:
    {\setlength{\abovedisplayskip}{3pt}
    \setlength{\belowdisplayskip}{3pt}
    \begin{equation*}
    \mathrm{MLA}(\mathsf{A}) \coloneqq \frac{1}{|\mathcal{E}|} \sum_{e\in \mathcal{E}}\left[\frac{acc(e_{D_\text{train}},D_\text{test}) - acc(e_S,D_\text{test})}{acc (e_{D_\text{train}},D_\text{test})
    }
    \right].
    \end{equation*}}
\end{definition}
A lower MLA score indicates a higher utility of synthetic data on the prediction task.

In addition to machine learning prediction, range/point queries are workhorses of statistical data analysis. 
\revision{For example, 3-way marginals were used in the NIST data synthesis challenge~\citep{nist18,icml19pgm,li2021dpsyn}. We extend it to support $k$-way marginals, as described below:}
\begin{definition} [Query Error]
    Consider a subset of $k$ attributes $a = \{a_1,...,a_k\}$ sampled from dataset $D$. For each attribute, if $a_i$ is categorical, a value $v_i$ is randomly chosen from its domain $\mathbb{R}(a_i)$, which forms the basis for a point query condition; for numerical attributes, two values $s_i$ and $d_i$ from $\mathbb{R}(a_i)$ are randomly sampled as the start and end points, to construct a range query condition. 
    The final query $c \in \mathcal{C}$ combines $k$ sub-queries and is executed on both real and synthetic data to obtain query frequency ratios $\mu_{c}^{D_\text{test}}$ and $\mu_{c}^S$. The query error of synthesizer $\mathsf{A}$ is: 
    {\setlength{\abovedisplayskip}{3pt}
    \setlength{\belowdisplayskip}{3pt}
    \begin{equation*}
        \mathrm{QueryError}(\mathsf{A}) \coloneqq 
        \frac{1}{|\mathcal{C}|} \sum_{c\in\mathcal{C}}
        \big[
        || \mu_{c}^{D_\text{test}} - \mu_{c}^S||_1 \big].
    \end{equation*}}
\end{definition}

%% file: 6_synmeter.tex
\begin{figure}[t]
    \centering
    \vspace{-3mm}
    \includegraphics[width=0.49\textwidth]{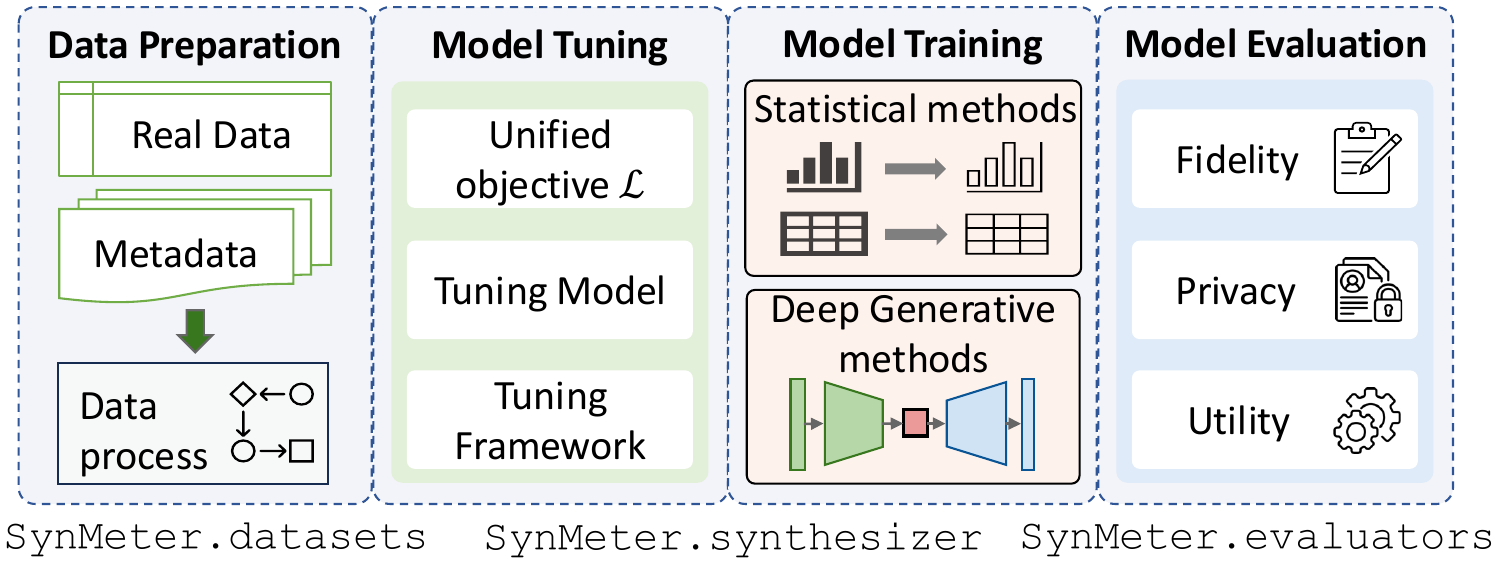}
    \vspace{-5mm}
    \caption{Overview of \mymethod.}
    \label{fig:framework}
\end{figure}

\section{A Systematic Evaluation Framework}
\label{sec:implements}

\mypara{Tuning Objective}
Most synthesizers~\cite{iclr23great,arxiv23realtabformer,iclr24tabsyn,nips19ctgan,usenix21privsyn} do not provide guidelines for hyperparameter tuning. Instead, default settings are commonly used for evaluations. This practice can lead to suboptimal results and biased comparisons.
To address this, we devise a simple tuning objective using proposed metrics to facilitate hyperparameter selection:
{\setlength{\abovedisplayskip}{3pt}
\setlength{\belowdisplayskip}{3pt}
\begin{equation}
\label{equ:tune_obj}
    \mathcal{L}(\mathsf{A}) = \alpha_1\mathrm{Fidelity} (\mathsf{A}) + \alpha_2\mathrm{MLA}(\mathsf{A}) + \alpha_3\mathrm{QueryError}(\mathsf{A}).
\end{equation}}The coefficients $\alpha_1$, $\alpha_2$, and $\alpha_3$ are assigned to three metrics.
Since smaller values indicate better performance for all proposed metrics, we conduct a grid search on synthesizers and select the best hyperparameters that minimize $\mathcal{L}$ for evaluation.
The privacy evaluation metric is excluded from model tuning, as we find that incorporating MDS results in negligible improvements for synthesizers while significantly increasing computational cost.
We show how to set the coefficients ($\alpha_1$, $\alpha_2$, $\alpha_3$) in Section~\ref{sec:exp_effective_metrics}.

\mypara{\mymethod}
We introduce a modular toolkit called \mymethod to assess synthesizers with the proposed evaluation metrics.  
As depicted in Figure~\ref{fig:framework}, \mymethod comprises four modules, and each module is implemented with an abstract interface for any synthesizer. 
We envisage that \mymethod can be used for the following purposes:
\begin{itemize}
    \item With eleven state-of-the-art synthesis algorithms implemented, it facilitates data owners to tune, train, and select different synthesizers for data publishing.
    \item It serves as a benchmark for data synthesis, providing systematic evaluation metrics for comparative studies.
    \item The modular architecture, connected via abstract interfaces, allows for easy integration of any synthesizers and enables adaptation into other domains (\eg genomic data synthesis~\cite{ndss22genomic}), by re-implementing module-specific processing functions.
\end{itemize}


%% file: 7_exp.tex
\section{Experiments}
\label{sec:exp}
Based on the proposed evaluation framework and metrics, we present a series of comprehensive experiments to answer the following question:
\begin{itemize}
    \item \textbf{RQ1:} How effective are our proposed privacy evaluation metric and tuning objective?
    \item \textbf{RQ2:} How do the various synthesizers perform under our assessment? What are the new findings?
    \item \textbf{RQ3:} Why do these methods work well (or not so well) on certain aspects? How can our metrics help with in-depth analysis? 
\end{itemize}

\input{7.1_setup}

\input{7.2_exp_metric}

\input{7.3_exp_eval}

\input{7.4_exp_ablation}

%% file: 7.1_setup.tex
\subsection{Experimental Setups}
\label{sec:exp_setup}

\mypara{Datasets}
\revision{We evaluate synthesizers on 12 real-world public datasets: Adult, Shoppers, Phishing, Magic, Faults, Bean, Obesity, Robot, Abalone, News, Insurance, and Wine.
These datasets vary in size, number of attributes, and underlying distributions.
We divide datasets into training and test with a ratio of 8:2, then split 20\% of the training dataset as the validation set, which is used for model tuning. 
Summary statistics for all datasets are provided in Table~\ref{tab:dataset_info}, and detailed descriptions can be found in Appendix~\ref{appendix:dataset}.
The results for the first six datasets are presented in this section, with the results for the remaining six datasets provided in the Appendix.}

\mypara{Data Synthesis Models}
We study a wide range of SOTA synthesizers, from statistical methods to deep generative models. We select them as they are either generally considered to perform best in practice~\citep{mckenna21mst,usenix21privsyn}, widely used~\citep{nips19ctgan,iclr18pate}, or recently emerged~\citep{icml23tabddpm,iclr23great,arxiv23tablediffusion}. These synthesizers can be categorized into two groups: heuristically private (HP) and differentially private (DP) synthesizers.

\noindent
\textit{HP Synthesizers} are developed without integrating DP:
\begin{itemize}
    \item \textbf{CTGAN}~\citep{nips19ctgan} is one of the most widely used HP synthesis algorithms. It utilizes generative adversarial networks to learn tabular data distributions. Training techniques like conditional generation and Wasserstein loss~\citep{nips17improvedwgan} are used.
    \item \textbf{TVAE}~\citep{nips19ctgan} is the state-of-the-art variational autoencoder for tabular data synthesizer, which uses mode-specific normalization to tackle the non-Gaussian problems of continuous distributions. 
    \item \textbf{TabDDPM}~\citep{icml23tabddpm} is the state-of-the-art diffusion model for data synthesis. It leverages the Gaussian diffusion process and the multinomial diffusion process to model continuous and discrete distributions, respectively.
    \item \textbf{TabSyn}~\citep{iclr24tabsyn} first trains an autoencoder to capture inter-column relationships and then employs a latent diffusion model to synthesize tabular data.
    \item \textbf{GReaT}~\cite{iclr23great} utilizes LLMs to convert records into textual representations and generate synthetic data through prompting.
    \item \textbf{REaLTabFormer}~\citep{arxiv23realtabformer} utilizes LLMs for data synthesis. It converts records to textual representations for LLM and generates synthetic data using fine-tuned GPT-2~\citep{openai2019llm}.
\end{itemize}

\noindent
\textit{DP Synthesizers} are either inherently designed with DP or are adaptations of HP models to offer provable privacy guarantees:
\begin{itemize}
    \item \textbf{MST}~\citep{mckenna21mst} is the state-of-the-art DP synthesizer, which uses probabilistic graphical models~\cite{icml19pgm} to learn the dependence of low-dimensional marginals. It won the NIST competition~\cite{nist18}. 
    Discrete binning is applied for numerical attributes.
    \item \textbf{PrivSyn}~\citep{usenix21privsyn} is a non-parametric DP synthesizer, which iteratively updates the synthetic dataset to make it match the target noise marginals. This method also shows strong performance in NIST competitions~\citep{nist18,nist20}. Discretization is also used for modeling numerical attributes.
    \item \textbf{PATE-GAN}~\citep{iclr18pategan} shares a similar architecture with CTGAN, but leverages the Private Aggregation of Teacher Ensembles (PATE)~\citep{iclr18pate} to offer DP guarantees.
    \item \textbf{TableDiffusion}~\citep{arxiv23tablediffusion} is a recently proposed diffusion model for data synthesis, which uses Differentially Private Stochastic Gradient Descent (DP-SGD) to enforce privacy.
    \item  \revision{\textbf{DP-2Stage}~\citep{tmlr25dp2stage} utilizes LLMs to first perform non-private template learning, capturing the structural patterns of the table, followed by differentially private fine-tuning using DP-SGD.}
\end{itemize}

All DP synthesizers can be adapted to the HP scenario either by using their HP counterparts\footnote{Although these paired models are quite different in the number of neural network layers, preprocessing, and learning strategies, they belong to the same type of generative model. Thus we call them ``counterparts''.} (\eg CTGAN for PATE-GAN, TabDDPM for TableDiffusion) or by setting the privacy budget to infinity (\eg MST and PrivSyn). 
However, some HP synthesizers (\eg TVAE) do not have corresponding DP variants. 
Thus, we only assess their performance within the context of HP models.

Note that our goal is not to benchmark all synthesizers but to focus on the best-known and broad spectrum of SOTA synthesizers.
We do not include workload-aware synthesizers like AIM~\citep{vldb22aim} since these approaches are optimized for answering pre-defined marginal queries, which are not considered for most synthesizers. 

\begin{table}[t]
    \centering
    \vspace{-3mm}
    \caption{Dataset Statistics. \# Num and \# Cat denote the number of numerical and categorical columns, respectively.}
    \vspace{-4mm}
    {\small \include{table/datasets_info}}
    \label{tab:dataset_info}
\end{table}

\mypara{Empirical Bound for Proposed Metrics}
We introduce additional baselines to better understand the performance of synthesizers on the proposed evaluation metrics:
\begin{itemize}
    \item \half randomly divides the real data into two equal parts: one serving as the training dataset $D$, and the other as the synthetic data $S$.  Since the two are drawn from the same distribution, this serves as an upper bound of fidelity.
    \item \hist is the simplified model of MST that only uses one-way marginals (\ie histograms) for data synthesis. Thus, in the synthetic dataset, all attributes are independent of each other.  We use \hist as the empirical lower bounds of fidelity. 
    \item \self uses a direct copy of the real data as synthetic data, establishing the worst privacy protection (lower bound). 
\end{itemize}

By the definition of MDS, the perfect privacy-preserving synthesizer would achieve a score of 0, which is the upper bound of privacy.
Having these lower and upper bounds for fidelity and privacy enables us to better interpret the scores of different synthesizers.  We do not include baselines for utility metrics (\ie machine learning affinity and query errors), since their scores are ratios that can naturally reflect the relative deviation from the ground truth.

\mypara{Implementation} 
We first tune the synthesizers with the proposed tuning objective. Then, synthetic data is generated by the trained synthesizer for evaluation. We test 20 times and report the mean and standard deviation as the final score for all evaluation metrics. 
The implementation details of the proposed metrics are in Appendix~\ref{appendix:implementation}, and the hyperparameter search spaces of all synthesis algorithms are in Appendix~\ref{appendix:hyperparameters}.



%% file: table/datasets_info.tex
\resizebox{0.48\textwidth}{!}{
\begin{tabular}{lccccccccc}
\toprule
\textbf{Dataset} & \# Train & \# Validation & \# Test & \# Num & \# Cat & Task type \\
\midrule
\textbf{Adult} & $20838$ & $5210$ & $6513$ & $6$ & $9$ & Binclass \\
\textbf{Shoppers} & $7891$ & $1973$ & $2466$ & $10$ & $8$  & Binclass \\
\textbf{Phishing} & $7075$ & $1769$ & $2211$ & $0$ & $31$  & Binclass \\
\textbf{Magic} &  $12172$ & $3044$ & $3804$ & $10$ & $1$  & Binclass \\
\textbf{Faults} & $1241$ & $311$ & $389$ & $24$ & $4$  & Multiclass(7) \\
\textbf{Bean} & $8710$ & $2178$ & $2723$ & $16$ & $1$  & Multiclass(7) \\
\textbf{Obesity} & $1350$ & $338$ & $423$ & $8$ & $9$  & Multiclass(7) \\
\textbf{Robot} & $3491$ & $873$ & $1092$ & $24$ & $1$ & Multiclass(4) \\
\textbf{Abalone} & $2672$ & $668$ & $836$ & $8$ & $1$  & Regression \\
\textbf{News}  & $25372$ & $6343$ & $7929$ & $46$ & $14$  & Regression \\
\textbf{Insurance}  & $856$ & $214$ & $268$ & $3$ & $4$ & Regression \\
\textbf{Wine}  & $3134$ & $784$ & $980$ & $12$ & $0$  & Regression \\
\bottomrule
\end{tabular}}

%% file: 7.2_exp_metric.tex
\begin{figure}[t]
    \centering
    \vspace{-3mm}
    \includegraphics[width=0.95\linewidth]{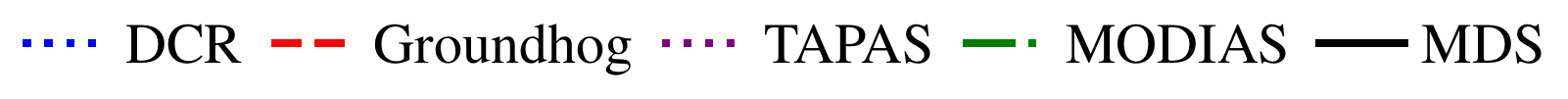}
    \subfigure[Evaluating DP synthesizer with privacy evaluation metrics.]
    {
    \includegraphics[width=0.45\linewidth]{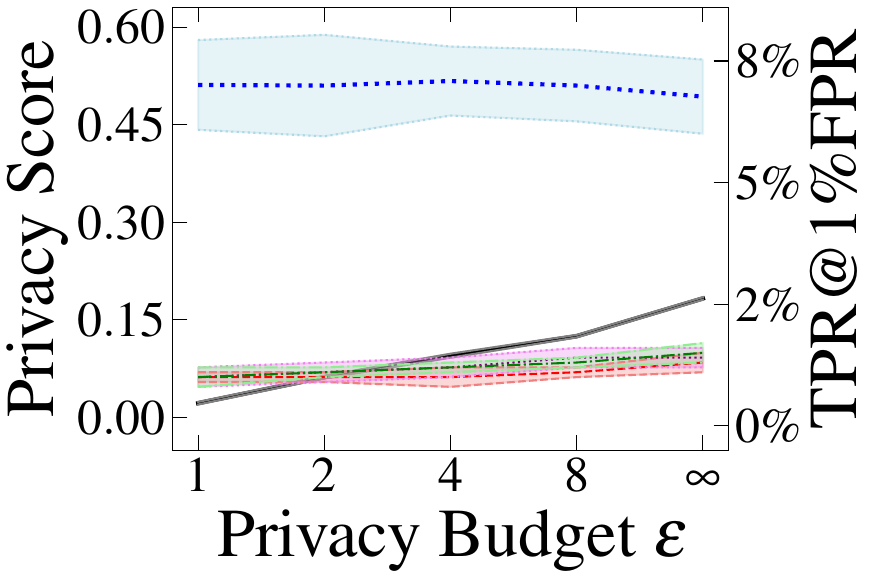}
    \label{fig:mds_dcr_mia_pategan}
    }
    \hfill
    \subfigure[Evaluating HP synthesizer with privacy evaluation metrics.]
    {
    \includegraphics[width=0.45\linewidth]{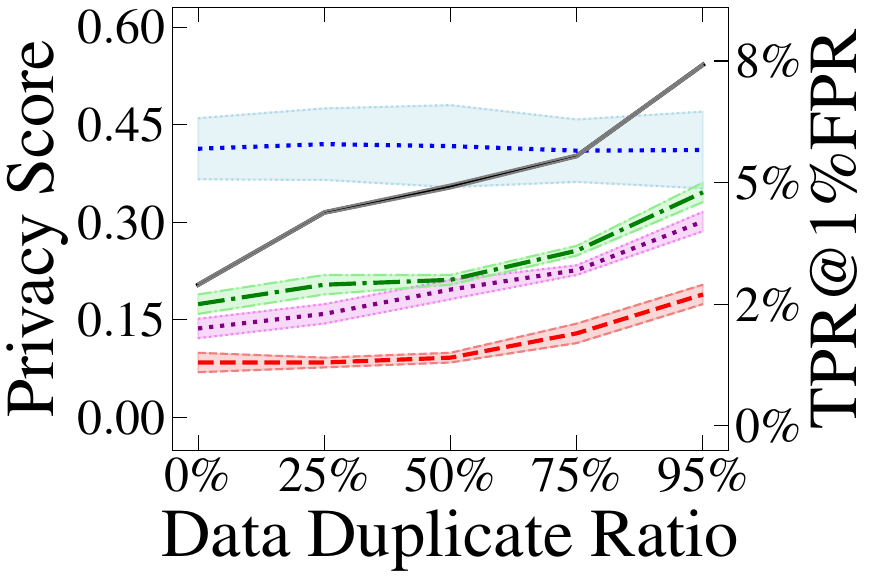}
    \label{fig:mds_dcr_mia_tabddpm}
    }
    \caption{Effectiveness evaluation of MDS on Adult dataset. DCR and MDS use the left y-axis (``Privacy Score'') whereas Groundhog, TAPAS, and MODIAS utilize the right y-axis (``TPR@1\%FPR'') for comparison. 
    Only MDS can distinguish different levels of privacy risks.}
    \label{fig:effectivness_mds}
\end{figure}

\begin{figure}[t]
    \centering
    \vspace{-3mm}
    \includegraphics[width=0.6\linewidth]{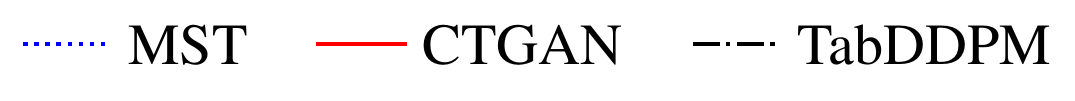}
    \subfigure[Impact of shadow models counts.]
    {
    \includegraphics[width=0.46\linewidth]{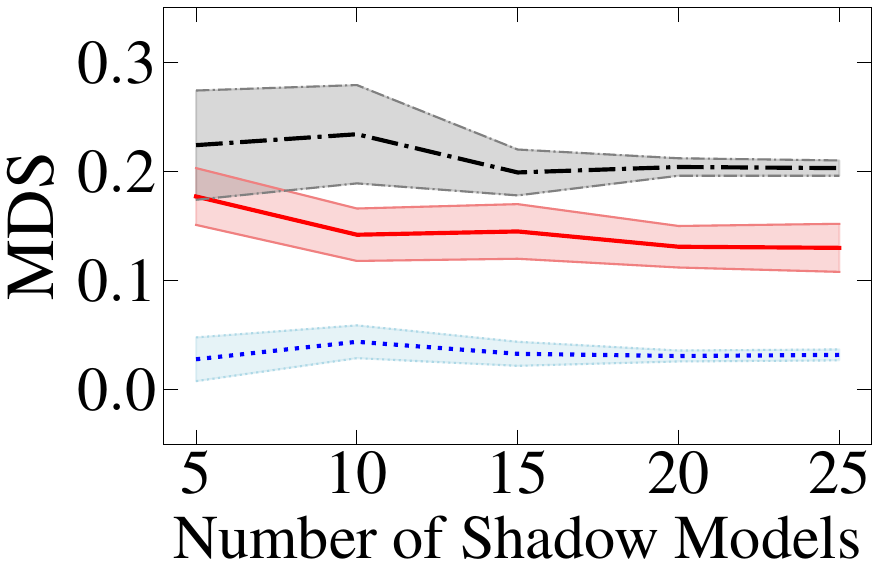}
    \label{fig:mds_m}
    }
    \hfill
    \subfigure[Impact of synthetic datasets counts.]
    {
    \includegraphics[width=0.46\linewidth]{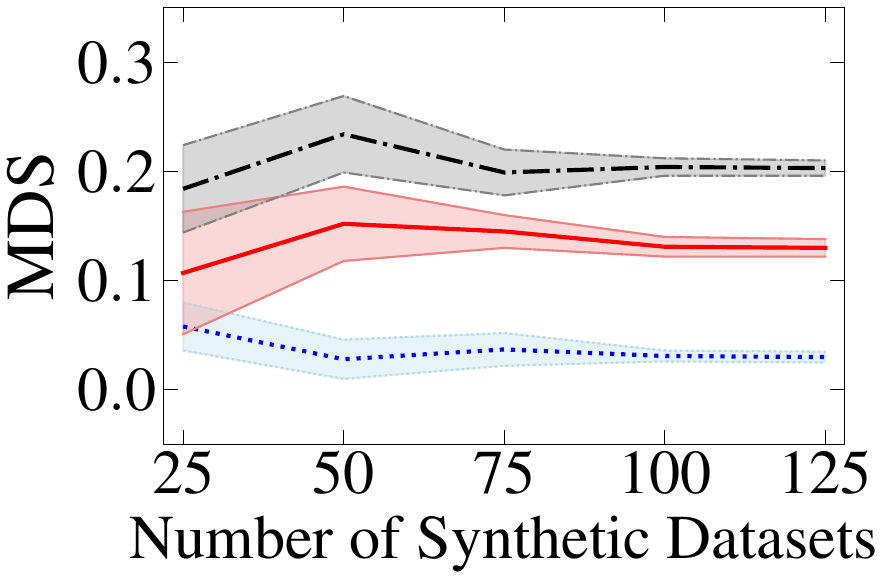}
    \label{fig:mds_n}
    }
    \vspace{-5mm}
    \caption{Stability evaluation of MDS on Adult dataset. We vary the number of shadow models and synthetic datasets used to calculate MDS. MDS can be reliably computed using 20 shadow models and 100 synthetic datasets.}
    \label{fig:stability_mds}
\end{figure}

\begin{figure*}[t]
    \centering
    \vspace{-1mm}
    \includegraphics[width=0.999\textwidth]{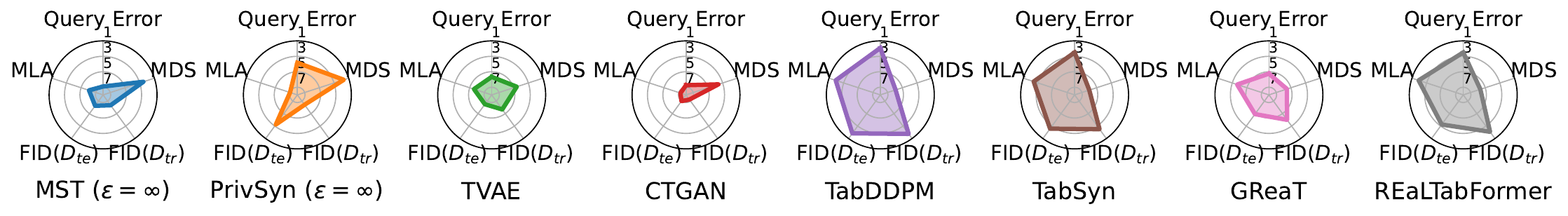}
    \vspace{-5mm}
    \caption{\revision{Average \textbf{ranking} comparison for HP synthesizers (outer means higher rank and better performance). Each vertex is the average rank of the method across 12 datasets, and each axis is one metric. ``FID($D_\text{tr}/D_\text{te}$)'' denotes the fidelity evaluated on the training/test dataset. ``MDS'' is the proposed privacy evaluation metric, and ``MLA'' and ``Query Error'' are utility metrics.}}
    \label{fig:ranking_nodp}
\end{figure*}

\begin{figure*}[t]
    \centering
    \vspace{-3mm}
    \includegraphics[width=0.999\textwidth]{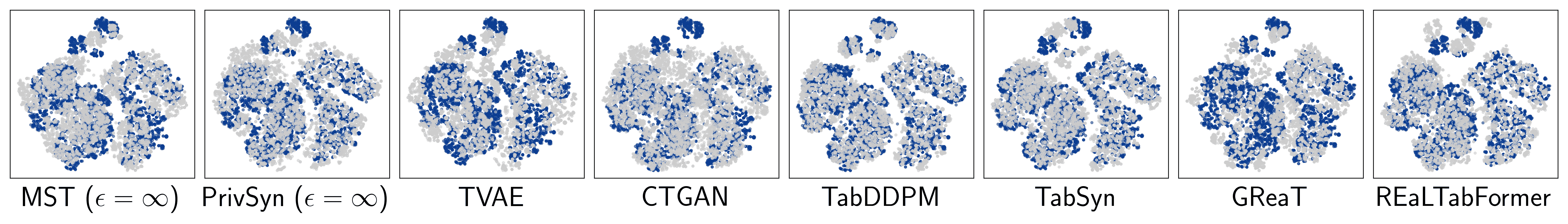}
    \vspace{-5mm}
    \caption{\revision{Visualization of HP synthesizers on Adult with t-SNE~\citep{tsne}. Real data are in blue and synthetic data are in grey.}}
    \label{fig:tsne_nodp}
\end{figure*}

\subsection{Effectiveness of MDS and Tuning Objective (RQ1)}
\label{sec:exp_effective_metrics}

\mypara{Effectiveness of MDS}
We compare MDS against the popular syntactic privacy evaluation metric DCR~\citep{acml21ctgabgan}, as well as three state-of-the-art MIAs against tabular data synthesis: Groundhog~\citep{usenix22tab_mia}, TAPAS~\citep{houssiau2022tapas}, and MODIAS~\citep{aistats23}. 
For DCR, we calculate the nearest distance of each synthetic record to real data, using the 5th percentile of the distance distribution as the privacy score. 
For MIAs, we follow~\citet{sp22lira} and use the true positive rate at 1\% false positive rate (TPR@1\%FPR) to measure the attack performance.
(We choose a false positive rate of 1\% because a lower false positive rate (e.g., 0.1\%) would result in a nearly zero true positive rate.)

We conduct two proof-of-concept experiments to evaluate the effectiveness of MDS. 
First, we train a DP synthesizer (\ie PATE-GAN) with varying levels of privacy protection by adjusting the privacy budget and then measure empirical privacy risk with different privacy evaluation metrics. 
Second, we train an HP synthesizer (\ie TabDDPM) with different duplication ratios while keeping the training data size unchanged.
Intuitively, a higher proportion of duplicate samples in the training set increases memorization, which in turn poses higher privacy risks~\citep{usenix23extracting}.

The results of both experiments are presented in Figure~\ref{fig:effectivness_mds}. 
DCR fails to distinguish between different levels of privacy risk in both scenarios and exhibits significant instability (indicated by large standard deviations). 
For MIAs, we observe an improvement in attack performance as the proportion of duplicates in the training set increases, especially for MODIAS. However, MIAs still struggle to capture privacy nuances with DP synthesizers. 
In contrast, MDS effectively detects privacy risks across all scenarios and demonstrates robustness as a reliable privacy evaluation metric, \revision{as evidenced by its low standard deviation}. 
We provide additional experiments to demonstrate the effectiveness of the proposed MDS and compare it with other evaluation metrics~\cite{meeus2023achilles}, as discussed in Appendix~\ref{appendix:comparsion_privacy_metrics}.


\begin{table}[t]
    \centering
    \vspace{-3mm}
    \caption{\revision{Running time of MDS across different synthesizers.}}
    \vspace{-4mm}
    {\small \include{table/effeciency_mds}}
    \label{tab:effeciency_mds}
\end{table}

\begin{table}[t]
    \centering
    \vspace{-3mm}
    \caption{Improvements (\%) with proposed tuning objective.}
    \vspace{-4mm}
    {\small \include{table/improv}}
    \label{tab:improv}
\end{table}

\begin{figure*}[t]
    \centering
    \vspace{-3mm}
    \includegraphics[width=0.8\textwidth]{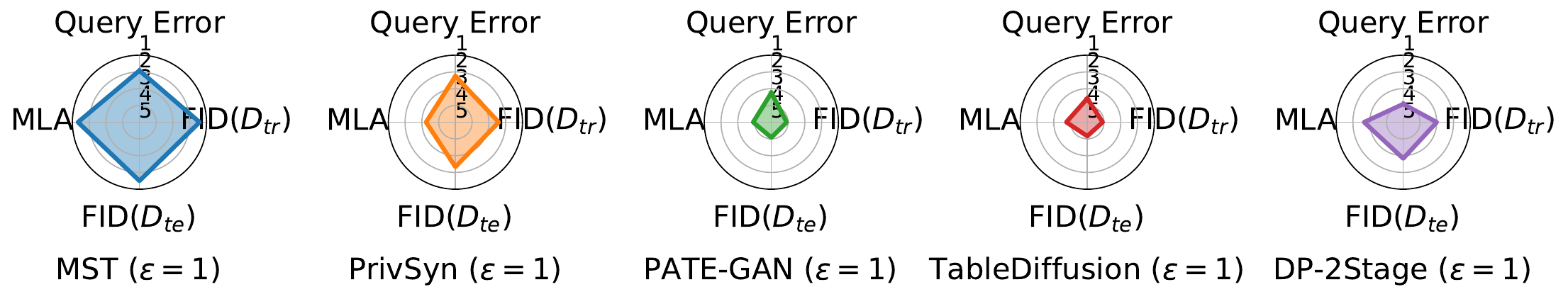}
    \vspace{-5mm}
    \caption{\revision{Average \textbf{ranking} comparison for DP synthesizers (outer means higher rank and better performance). All methods offer provable privacy guarantees, so the privacy axis is omitted.}}
    \label{fig:ranking_dp}
\end{figure*}

\begin{figure*}[t]
    \centering
    \vspace{-3mm}
    \includegraphics[width=0.73\textwidth]{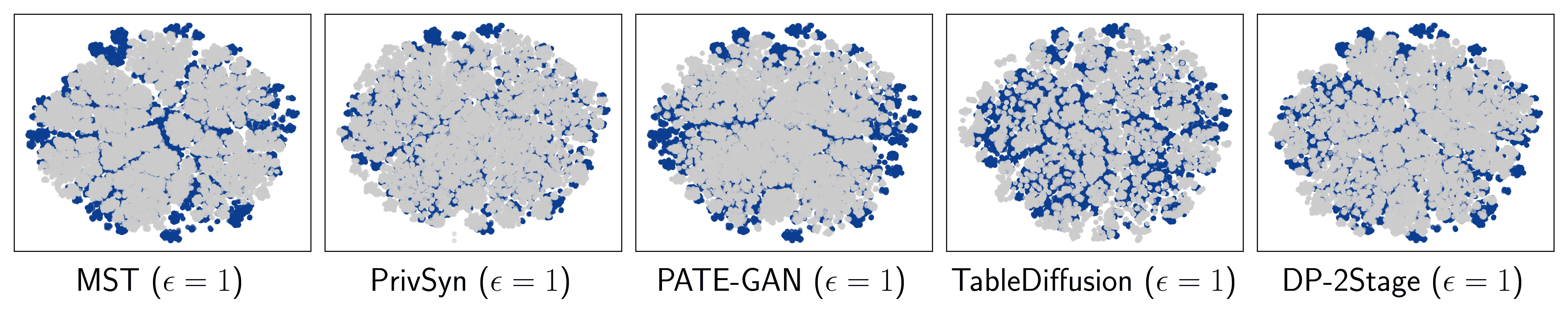}
    \vspace{-5mm}
    \caption{\revision{Visualization of DP synthesizers on Adult with t-SNE. Real data are in blue and synthetic data are in grey.}}
    \label{fig:tsne_dp}
\end{figure*}

\mypara{Stability and Efficiency of MDS}
We validate the stability of MDS by varying the number of shadow models and synthetic datasets. Specifically, we compute the privacy scores for three synthesizers using different quantities of shadow models and synthetic datasets, recording the mean and variance of the results, as depicted in Figure~\ref{fig:stability_mds}. Our results indicate that the variance of MDS decreases rapidly as the number of shadow models and synthetic datasets increases, with stable results achieved using 20 shadow models and 100 synthetic datasets. 
\revision{We also evaluate the efficiency of the proposed privacy evaluation metric using a cluster with an A100 GPU. As shown in Table~\cref{tab:effeciency_mds}, MDS can be efficiently computed for most synthesizers. While REaLTabFormer requires more time, the overall computational cost remains practical.}

\mypara{Effectiveness of Tuning Objective}
Although the metrics in Equation (\ref{equ:tune_obj}) are based on different measurements, empirically we observe that their values consistently fall within the same range. 
Consequently, 
In our experiments, we set all three coefficients to $1/3$, as this configuration significantly improves the quality of synthetic data, as shown in Table~\ref{tab:improv}.
Interestingly, the tuning phase affects two types of synthesizers differently: statistical methods gain more utility than fidelity, while deep generative models show the opposite trend.
Notably, the tuning phase proves especially beneficial for diffusion models (\ie TabDDPM and TableDiffusion), with notable improvements in both fidelity and utility.
Additional experiments of the tuning objective are provided in Appendix~\ref{appendix:model_tuning}.


%% file: table/effeciency_mds.tex
\begin{tabular}{ccc}
    \toprule[0.8pt]
    \textbf{Synthesizer} & \textbf{Training (hrs)} & \textbf{Sampling (hrs)}\\
    \midrule
    MST & $0.1$ & $0.02$ \\
    PrivSyn & - & $0.18$ \\
    TVAE & $0.2$ & $0.08$ \\
    CTGAN & $0.2$ & $0.08$ \\
    PATE-GAN & $0.23$ & $0.08$ \\
    TabDDPM & $0.25$ & $0.09$ \\
    TabSyn & $0.22$ & $0.08$ \\
    GReaT & $7.89$ & $1.83$ \\
    REaLTabFormer & $2.57$ & $0.37$ \\
    TableDiffusion & $1.28$ & $0.10$ \\
    DP-2Stage & $4.28$ & $0.50$ \\
    \bottomrule[1.0pt]
\end{tabular}

%% file: table/improv.tex
\resizebox{0.4\textwidth}{!}{\begin{tabular}{cccccc}
    \toprule
    \multicolumn{1}{c}{\multirow{2}{*}{\textbf{Synthesizer}}} & \multicolumn{2}{c}{\textbf{Fidelity} $\uparrow$} & \multicolumn{2}{c}{\textbf{Utility $\uparrow$}} \\
    \cmidrule(lr){2-3}  \cmidrule(lr){4-5}
    & $D_\text{Train}$ & $D_\text{Test}$ & MLA & Query Error \\
    \midrule
    MST & $0.33$ & $0.34$ & $17.35$ & $3.39$ \\
    PrivSyn & $1.60$ & $2.92$ & $12.08$ & $1.12$ \\
    TVAE & $1.06$ & $0.67$ & $5.29$ & $2.67$ \\
    CTGAN & $9.87$ & $9.60$ & $0.57$ & $8.63$ \\
    PATE-GAN & $6.27$ & $8.48$ & $0.75$ & $7.04$ \\
    TabDDPM & $13.62$ & $13.65$ & $13.67$ & $11.95$\\
    TabSyn & $2.16$ & $3.84$ & $2.74$ & $2.95$\\
    GReaT & $3.84$ & $9.21$ & $1.14$ & $1.77$ \\
    REaLTabFormer  & $2.31$ & $4.48$ & $3.73$ & $5.29$ \\
    TableDiffusion & $11.34$ & $10.95$ & $8.32$ & $7.86$ \\
    DP-2Stage & $5.93$ & $6.24$ & $3.76$ & $3.05$ \\
    \bottomrule
\end{tabular}}

%% file: 7.3_exp_eval.tex
\begin{table*}[t]
    \centering
    \vspace{-3mm}
    \caption{Fidelity evaluation (lower score means better fidelity) of synthesizers on training data $D_\text{train}$ of the first six datasets (the results for the last six datasets are in Table~\ref{tab:train_fidelity_2}). The privacy budget $\epsilon$ of HP synthesizers is  $\infty$ (the top part). \half and \hist are the baselines that serve as the empirical upper/lower bounds of the fidelity for HP synthesizers. The best result is in bold.}
    \vspace{-3mm}
    \label{tab:train_fidelity}
    {\include{table/train_fidelity}}
\end{table*}

\begin{table*}[h]
    \centering
    \vspace{-3mm}
    \caption{Fidelity evaluation of synthesizers on test data $D_\text{test}$ of the first six datasets (the results for the last six are in Table~\ref{tab:test_fidelity_2}). 
    }
    \vspace{-3mm}
    \label{tab:test_fidelity}
    {\include{table/test_fidelity}}
\end{table*}

\subsection{Overall Evaluation (RQ2)}\label{sec:exp_overall}

\mypara{Overview}
Figure~\ref{fig:ranking_nodp} and Figure~\ref{fig:ranking_dp} report the overview ranking results for HP and DP synthesizers, respectively. 
For HP synthesizers, TabDDPM and REaLTabFormer exhibit superior fidelity and utility, albeit at the expense of compromising privacy. 
Statistical methods like PrivSyn achieve good fidelity while offering impressive privacy protection.  
Conversely, CTGAN, the most popular HP synthesizer, shows the least satisfactory results in synthetic data quality.
For DP synthesizers, statistical methods remain effective in both fidelity and utility. 
The performance of deep generative models drops significantly to satisfy differential privacy. Even the strongest model (\ie TableDiffusion) underperforms statistical approaches by a large margin, which starkly contrasts with its performance in the HP context, indicating a pronounced impact of privacy constraints on deep generative models.
We also visualize the synthetic and real data using t-SNE~\citep{tsne}, as shown in Figure~\ref{fig:tsne_nodp} and Figure~\ref{fig:tsne_dp}. 
The visualizations generally reveal substantial overlap between synthetic and real data points for all synthesizers.



\mypara{Fidelity Evaluation} 
Fidelity is evaluated by applying the Wasserstein distance to both the training dataset $D_\text{train}$ (Table~\ref{tab:train_fidelity} and Table~\ref{tab:train_fidelity_2}) and the test dataset $D_\text{test}$ (Table~\ref{tab:test_fidelity} and Table~\ref{tab:test_fidelity_2}). 
The results show that TabDDPM, TabSyn, and REaLTabFormer achieve near upper-bound fidelity, while statistical methods such as MST and PrivSyn excel among DP synthesizers. 
Notably, all deep generative models experience a significant drop in fidelity when achieving differential privacy, whereas statistical methods maintain consistent performance. 
This stability is further emphasized when comparing training and test data, where statistical methods show a smaller performance drop than deep generative models.

\begin{table*}[ht]
    \centering
    \vspace{-3mm}
    \caption{\revision{Privacy evaluation (lower score means better empirical privacy protection) of synthesizers on the first six datasets (the results for the last six datasets are in Table~\ref{tab:privacy_2}). \self is the baseline that serves as the empirical lower bound of MDS.
    }}
    \label{tab:privacy}
    \vspace{-3mm}
    {\include{table/privacy}}
\end{table*}

\begin{table*}[ht]
    \centering
    \vspace{-3mm}
    \caption{Utility evaluation (\ie MLA, lower score means better utility) of synthesizers on the first six datasets (the results for the last six datasets are in Table~\ref{tab:mla_2}). The privacy budget $\epsilon$ of HP synthesizers is set as $\infty$ (the top part).
    }
    \label{tab:mla}
    \vspace{-3mm}
    {\include{table/mla}}
\end{table*}

\mypara{Privacy Evaluation}
Table~\ref{tab:privacy} and Table~\ref{tab:privacy_2} show the privacy assessment results for synthesizers. 
For HP synthesizers, statistical methods like MST and PrivSyn also show notable empirical privacy protections. 
However, the unsatisfactory results of strong synthesizers like TabDDPM reveal their vulnerability to membership disclosure. 
\revision{For DP synthesizers, we observe that all DP synthesizers demonstrate strong empirical privacy guarantees, with MDS values approaching the theoretical upper bound of 0.}

\begin{table*}[t]
    \centering
    \vspace{-3mm}
    \caption{Utility evaluation (\ie query error) of synthesizers on the first six datasets (the results for the last six are in Table~\ref{tab:query_2}). 
    }
    \label{tab:query}
    \vspace{-3mm}
    {\include{table/query}}
\end{table*}

\mypara{Utility Evaluation}
The utility of data synthesis is assessed by performing downstream tasks on the synthetic datasets and measuring their performance using the proposed metrics, as shown in 
Table~\ref{tab:mla}, Table~\ref{tab:query} Table~\ref{tab:mla_2}, and Table~\ref{tab:query_2}.
For machine learning tasks, TabDDPM excels among HP synthesizers, contributing to its class-conditional framework that learns label dependencies during its training process. However, this advantage diminishes when adding random noise to ensure privacy, where MST takes the lead with its robust and superior performance. The outcomes for range (point) query tasks echo the results of fidelity evaluation, where TabDDPM shows superior performance in HP settings, and statistical methods (\eg MST) can surpass other methods under DP constraints.

%% file: table/train_fidelity.tex
\resizebox{0.75\textwidth}{!}{
\begin{tabular}{lccccccc}
\toprule[0.8pt]
 & \textbf{Adult} & \textbf{Shoppers} & \textbf{Phishing} & \textbf{Magic} & \textbf{Faults} & \textbf{Bean} \\
\midrule
MST & $0.186\scriptscriptstyle \pm \scriptstyle .010$ & $0.092\scriptscriptstyle \pm \scriptstyle .002$ & $0.019\scriptscriptstyle \pm \scriptstyle .001$ & $0.037\scriptscriptstyle \pm \scriptstyle .002$ & $0.056\scriptscriptstyle \pm \scriptstyle .002$ & $0.040\scriptscriptstyle \pm \scriptstyle .002$  \\
PrivSyn & $0.024\scriptscriptstyle \pm \scriptstyle .001$ & $0.030\scriptscriptstyle \pm \scriptstyle .001$ & $\mathbf{0.010\scriptscriptstyle \pm \scriptstyle .001}\mathbf{}$ & $0.015\scriptscriptstyle \pm \scriptstyle .003$ & $0.064\scriptscriptstyle \pm \scriptstyle .006$ & $0.035\scriptscriptstyle \pm \scriptstyle .002$  \\
TVAE & $0.085\scriptscriptstyle \pm \scriptstyle .002$ & $0.156\scriptscriptstyle \pm \scriptstyle .001$ & $0.024\scriptscriptstyle \pm \scriptstyle .001$ & $0.021\scriptscriptstyle \pm \scriptstyle .003$ & $0.055\scriptscriptstyle \pm \scriptstyle .007$ & $0.047\scriptscriptstyle \pm \scriptstyle .006$  \\
CTGAN & $0.059\scriptscriptstyle \pm \scriptstyle .001$ & $0.062\scriptscriptstyle \pm \scriptstyle .001$ & $0.062\scriptscriptstyle \pm \scriptstyle .002$ & $0.157\scriptscriptstyle \pm \scriptstyle .006$ & $0.133\scriptscriptstyle \pm \scriptstyle .004$ & $0.139\scriptscriptstyle \pm \scriptstyle .005$  \\
TabDDPM & $\mathbf{0.020\scriptscriptstyle \pm \scriptstyle .001}$ & $\mathbf{0.022\scriptscriptstyle \pm \scriptstyle .001}$ & $0.015\scriptscriptstyle \pm \scriptstyle .001$ & $\mathbf{0.011\scriptscriptstyle \pm \scriptstyle .003}$ & $\mathbf{0.026\scriptscriptstyle \pm \scriptstyle .002}$ & $\mathbf{0.015\scriptscriptstyle \pm \scriptstyle .002}$ \\
TabSyn & $0.025\scriptscriptstyle \pm \scriptstyle .003$ & $0.030\scriptscriptstyle \pm \scriptstyle .005$ & $0.018\scriptscriptstyle \pm \scriptstyle .003$ & $0.012\scriptscriptstyle \pm \scriptstyle .004$ & $0.034\scriptscriptstyle \pm \scriptstyle .003$ & $0.033\scriptscriptstyle \pm \scriptstyle .008$ &  \\
GReaT & $0.050\scriptscriptstyle \pm \scriptstyle .002$ & $0.049\scriptscriptstyle \pm \scriptstyle .003$ & $0.076\scriptscriptstyle \pm \scriptstyle .002$ & $0.037\scriptscriptstyle \pm \scriptstyle .003$ & $0.050\scriptscriptstyle \pm \scriptstyle .006$ & $0.020\scriptscriptstyle \pm \scriptstyle .001$  \\
REaLTabFormer & $0.022\scriptscriptstyle \pm \scriptstyle .002$ & $0.024\scriptscriptstyle \pm \scriptstyle .003$ & $0.012\scriptscriptstyle \pm \scriptstyle .001$ & $0.045\scriptscriptstyle \pm \scriptstyle .005$ & $0.054\scriptscriptstyle \pm \scriptstyle .005$ & $0.035\scriptscriptstyle \pm \scriptstyle .007$ \\
\midrule
MST ($\epsilon=1$) & $0.198\scriptscriptstyle \pm \scriptstyle .013$ & $0.103\scriptscriptstyle \pm \scriptstyle .002$ & $\mathbf{0.023\scriptscriptstyle \pm \scriptstyle .001}$ & $\mathbf{0.042\scriptscriptstyle \pm \scriptstyle .003}$ & $\mathbf{0.086\scriptscriptstyle \pm \scriptstyle .003}$ & $\mathbf{0.048\scriptscriptstyle \pm \scriptstyle .004}$   \\
PrivSyn ($\epsilon=1$) & $\mathbf{0.045\scriptscriptstyle \pm \scriptstyle .002}$ & $\mathbf{0.077\scriptscriptstyle \pm \scriptstyle .005}$ & $0.033\scriptscriptstyle \pm \scriptstyle .002$ & $0.052\scriptscriptstyle \pm \scriptstyle .003$ & $0.228\scriptscriptstyle \pm \scriptstyle .007$ & $0.142\scriptscriptstyle \pm \scriptstyle .007$  \\
PATE-GAN ($\epsilon=1$) & $0.139\scriptscriptstyle \pm \scriptstyle .001$ & $0.176\scriptscriptstyle \pm \scriptstyle .002$ & $0.173\scriptscriptstyle \pm \scriptstyle .002$ & $0.153\scriptscriptstyle \pm \scriptstyle .005$ & $0.204\scriptscriptstyle \pm \scriptstyle .003$ & $0.520\scriptscriptstyle \pm \scriptstyle .006$ \\
TableDiffusion ($\epsilon=1$) & $0.180\scriptscriptstyle \pm \scriptstyle .002$ & $0.209\scriptscriptstyle \pm \scriptstyle .002$ & $0.123\scriptscriptstyle \pm \scriptstyle .002$ & $0.132\scriptscriptstyle \pm \scriptstyle .003$ & $0.369\scriptscriptstyle \pm \scriptstyle .002$ & $0.148\scriptscriptstyle \pm \scriptstyle .005$  \\
DP-2Stage ($\epsilon=1$) & $0.112\scriptscriptstyle \pm \scriptstyle .002$ & $0.192\scriptscriptstyle \pm \scriptstyle .007$ & $0.085\scriptscriptstyle \pm \scriptstyle .008$ & $0.059\scriptscriptstyle \pm \scriptstyle .007$ & $0.131\scriptscriptstyle \pm \scriptstyle .004$ & $0.083\scriptscriptstyle \pm \scriptstyle .006$  \\
\midrule
\midrule
\half (upper bound) & $0.020\scriptscriptstyle \pm \scriptstyle .002$ & $0.018\scriptscriptstyle \pm \scriptstyle .001$ & $0.010\scriptscriptstyle \pm \scriptstyle .002$ & $0.011\scriptscriptstyle \pm \scriptstyle .004$ & $0.017\scriptscriptstyle \pm \scriptstyle .002$ & $0.015\scriptscriptstyle \pm \scriptstyle .004$  \\
\hist (lower bound) & $0.213\scriptscriptstyle \pm \scriptstyle .013$ & $0.101\scriptscriptstyle \pm \scriptstyle .003$ & $0.027\scriptscriptstyle \pm \scriptstyle .001$ & $0.051\scriptscriptstyle \pm \scriptstyle .003$ & $0.081\scriptscriptstyle \pm \scriptstyle .002$ & $0.087\scriptscriptstyle \pm \scriptstyle .002$  \\
\bottomrule[0.8pt]
\end{tabular}}

%% file: table/test_fidelity.tex
\resizebox{0.75\textwidth}{!}{
\begin{tabular}{lcccccccccccc}
\toprule[0.8pt]
 & \textbf{Adult} & \textbf{Shoppers} & \textbf{Phishing} & \textbf{Magic} & \textbf{Faults} & \textbf{Bean} \\
\midrule
MST & $0.172\scriptscriptstyle \pm \scriptstyle .004$ & $0.098\scriptscriptstyle \pm \scriptstyle .002$ & $0.026\scriptscriptstyle \pm \scriptstyle .001$ & $0.039\scriptscriptstyle \pm \scriptstyle .002$ & $0.089\scriptscriptstyle \pm \scriptstyle .006$ & $0.044\scriptscriptstyle \pm \scriptstyle .003$  \\
PrivSyn & $0.025\scriptscriptstyle \pm \scriptstyle .001$ & $0.041\scriptscriptstyle \pm \scriptstyle .003$ & $\mathbf{0.017\scriptscriptstyle \pm \scriptstyle .002}$ & $0.015\scriptscriptstyle \pm \scriptstyle .002$ & $0.079\scriptscriptstyle \pm \scriptstyle .007$ & $0.037\scriptscriptstyle \pm \scriptstyle .003$  \\
TVAE & $0.086\scriptscriptstyle \pm \scriptstyle .002$ & $0.154\scriptscriptstyle \pm \scriptstyle .002$ & $0.028\scriptscriptstyle \pm \scriptstyle .002$ & $0.020\scriptscriptstyle \pm \scriptstyle .003$ & $0.081\scriptscriptstyle \pm \scriptstyle .016$ & $0.050\scriptscriptstyle \pm \scriptstyle .004$  \\
CTGAN & $0.061\scriptscriptstyle \pm \scriptstyle .003$ & $0.061\scriptscriptstyle \pm \scriptstyle .002$ & $0.069\scriptscriptstyle \pm \scriptstyle .001$ & $0.150\scriptscriptstyle \pm \scriptstyle .004$ & $0.133\scriptscriptstyle \pm \scriptstyle .007$ & $0.139\scriptscriptstyle \pm \scriptstyle .005$  \\
TabDDPM & $\mathbf{0.021\scriptscriptstyle \pm \scriptstyle .001}$ & $0.031\scriptscriptstyle \pm \scriptstyle .001$ & $0.019\scriptscriptstyle \pm \scriptstyle .001$ & $\mathbf{0.012\scriptscriptstyle \pm \scriptstyle .002}$ & $0.058\scriptscriptstyle \pm \scriptstyle .008$ & $\mathbf{0.016\scriptscriptstyle \pm \scriptstyle .003}$  \\
TabSyn & $0.028\scriptscriptstyle \pm \scriptstyle .002$ & $0.035\scriptscriptstyle \pm \scriptstyle .001$ & $0.023\scriptscriptstyle \pm \scriptstyle .001$ & $0.014\scriptscriptstyle \pm \scriptstyle .002$ & $\mathbf{0.057\scriptscriptstyle \pm \scriptstyle .012}$ & $0.033\scriptscriptstyle \pm \scriptstyle .007$ \\
GReaT & $0.052\scriptscriptstyle \pm \scriptstyle .002$ & $0.056\scriptscriptstyle \pm \scriptstyle .004$ & $0.072\scriptscriptstyle \pm \scriptstyle .002$ & $0.039\scriptscriptstyle \pm \scriptstyle .003$ & $0.063\scriptscriptstyle \pm \scriptstyle .007$ & $0.021\scriptscriptstyle \pm \scriptstyle .004$  \\
REaLTabFormer & $\mathbf{0.021\scriptscriptstyle \pm \scriptstyle .001}$ & $\mathbf{0.030\scriptscriptstyle \pm \scriptstyle .003}$ & $0.018\scriptscriptstyle \pm \scriptstyle .004$ & $0.046\scriptscriptstyle \pm \scriptstyle .003$ & $0.075\scriptscriptstyle \pm \scriptstyle .005$ & $0.028\scriptscriptstyle \pm \scriptstyle .004$ \\
\midrule
MST ($\epsilon=1$) & $0.179\scriptscriptstyle \pm \scriptstyle .004$ & $0.103\scriptscriptstyle \pm \scriptstyle .001$ & $\mathbf{0.028\scriptscriptstyle \pm \scriptstyle .001}$ & $0.042\scriptscriptstyle \pm \scriptstyle .004$ & $\mathbf{0.112\scriptscriptstyle \pm \scriptstyle .005}$ & $\mathbf{0.048\scriptscriptstyle \pm \scriptstyle .003}$  \\
PrivSyn ($\epsilon=1$) & $\mathbf{0.049\scriptscriptstyle \pm \scriptstyle .002}$ & $\mathbf{0.084\scriptscriptstyle \pm \scriptstyle .002}$ & $0.030\scriptscriptstyle \pm \scriptstyle .003$ & $\mathbf{0.031\scriptscriptstyle \pm \scriptstyle .003}$ & $0.236\scriptscriptstyle \pm \scriptstyle .017$ & $0.128\scriptscriptstyle \pm \scriptstyle .010$  \\
PATE-GAN ($\epsilon=1$) & $0.139\scriptscriptstyle \pm \scriptstyle .002$ & $0.171\scriptscriptstyle \pm \scriptstyle .002$ & $0.173\scriptscriptstyle \pm \scriptstyle .002$ & $0.155\scriptscriptstyle \pm \scriptstyle .005$ & $0.215\scriptscriptstyle \pm \scriptstyle .004$ & $0.523\scriptscriptstyle \pm \scriptstyle .004$  \\
TableDiffusion ($\epsilon=1$) & $0.179\scriptscriptstyle \pm \scriptstyle .002$ & $0.210\scriptscriptstyle \pm \scriptstyle .002$ & $0.121\scriptscriptstyle \pm \scriptstyle .002$ & $0.132\scriptscriptstyle \pm \scriptstyle .005$ & $0.390\scriptscriptstyle \pm \scriptstyle .004$ & $0.149\scriptscriptstyle \pm \scriptstyle .004$  \\
DP-2Stage ($\epsilon=1$) & $0.120\scriptscriptstyle \pm \scriptstyle .003$ & $0.172\scriptscriptstyle \pm \scriptstyle .005$ & $0.107\scriptscriptstyle \pm \scriptstyle .004$ & $0.076\scriptscriptstyle \pm \scriptstyle .003$ & $0.192\scriptscriptstyle \pm \scriptstyle .002$ & $0.073\scriptscriptstyle \pm \scriptstyle .002$  \\
\midrule
\midrule
\half (upper bound) & $0.022\scriptscriptstyle \pm \scriptstyle .002$ & $0.023\scriptscriptstyle \pm \scriptstyle .002$ & $0.016\scriptscriptstyle \pm \scriptstyle .003$ & $0.011\scriptscriptstyle \pm \scriptstyle .003$ & $0.042\scriptscriptstyle \pm \scriptstyle .005$ & $0.015\scriptscriptstyle \pm \scriptstyle .003$  \\
\hist (lower bound) & $0.199\scriptscriptstyle \pm \scriptstyle .017$ & $0.101\scriptscriptstyle \pm \scriptstyle .001$ & $0.030\scriptscriptstyle \pm \scriptstyle .001$ & $0.048\scriptscriptstyle \pm \scriptstyle .002$ & $0.113\scriptscriptstyle \pm \scriptstyle .006$ & $0.080\scriptscriptstyle \pm \scriptstyle .003$  \\
\bottomrule[1.0pt]
\end{tabular}}

%% file: table/privacy.tex
\resizebox{0.75\textwidth}{!}{
\begin{tabular}{lcccccccccccc}
\toprule[0.8pt]
 & \textbf{Adult} & \textbf{Shoppers} & \textbf{Phishing} & \textbf{Magic} & \textbf{Faults} & \textbf{Bean}\\
\midrule
MST 
& $\mathbf{0.031 \scriptscriptstyle \pm \scriptstyle .001}$ 
& $0.012 \scriptscriptstyle \pm \scriptstyle .002$ 
& $0.038 \scriptscriptstyle \pm \scriptstyle .003$ 
& $0.008 \scriptscriptstyle \pm \scriptstyle .001$ 
& $ 0.030 \scriptscriptstyle \pm \scriptstyle .002$ 
& $ 0.015 \scriptscriptstyle \pm \scriptstyle .003$  \\
PrivSyn 
& $0.046 \scriptscriptstyle \pm \scriptstyle .002$ 
& $\mathbf{0.005 \scriptscriptstyle \pm \scriptstyle .001}$ 
& $0.017 \scriptscriptstyle \pm \scriptstyle .003$ 
& $\mathbf{0.005 \scriptscriptstyle \pm \scriptstyle .002}$ 
& $\mathbf{0.004 \scriptscriptstyle \pm \scriptstyle .001}$ 
& $\mathbf{0.006 \scriptscriptstyle \pm \scriptstyle .003}$ 
\\
TVAE 
& $0.192 \scriptscriptstyle \pm \scriptstyle .003$ 
& $0.050 \scriptscriptstyle \pm \scriptstyle .002$ 
& $\mathbf{0.016 \scriptscriptstyle \pm \scriptstyle .001}$ 
& $0.016 \scriptscriptstyle \pm \scriptstyle .005$ 
& $0.037 \scriptscriptstyle \pm \scriptstyle .002$ 
& $0.029 \scriptscriptstyle \pm \scriptstyle .001$ 
 \\
CTGAN 
& $0.131 \scriptscriptstyle \pm \scriptstyle .002$ 
& $0.018 \scriptscriptstyle \pm \scriptstyle .003$ 
& $0.125 \scriptscriptstyle \pm \scriptstyle .001$ 
& $0.012 \scriptscriptstyle \pm \scriptstyle .003$ 
& $0.011 \scriptscriptstyle \pm \scriptstyle .003$ 
& $0.028 \scriptscriptstyle \pm \scriptstyle .001$ 
\\
TabDDPM 
& $0.204 \scriptscriptstyle \pm \scriptstyle .001$ 
& $0.019 \scriptscriptstyle \pm \scriptstyle .002$ 
& $0.082 \scriptscriptstyle \pm \scriptstyle .003$ 
& $0.015 \scriptscriptstyle \pm \scriptstyle .001$ 
& $0.092 \scriptscriptstyle \pm \scriptstyle .002$ 
& $0.020 \scriptscriptstyle \pm \scriptstyle .003$ 
\\
TabSyn 
& $0.202 \scriptscriptstyle \pm \scriptstyle .001$ 
& $0.017 \scriptscriptstyle \pm \scriptstyle .003$ 
& $0.088 \scriptscriptstyle \pm \scriptstyle .002$ 
& $0.029 \scriptscriptstyle \pm \scriptstyle .001$
& $0.100 \scriptscriptstyle \pm \scriptstyle .003 $
& $0.021 \scriptscriptstyle \pm \scriptstyle .003$  
\\
GReaT 
& $0.199 \scriptscriptstyle \pm \scriptstyle .002$ 
& $0.044 \scriptscriptstyle \pm \scriptstyle .003$ 
& $0.091 \scriptscriptstyle \pm \scriptstyle .001$ 
& $0.011 \scriptscriptstyle \pm \scriptstyle .002$ 
& $0.099 \scriptscriptstyle \pm \scriptstyle .003$ 
& $0.016 \scriptscriptstyle \pm \scriptstyle .004$  
\\
REaLTabFormer 
& $0.234 \scriptscriptstyle \pm \scriptstyle .001$
& $0.047 \scriptscriptstyle \pm \scriptstyle .002$
& $0.084 \scriptscriptstyle \pm \scriptstyle .003$ 
& $0.011 \scriptscriptstyle \pm \scriptstyle .002$ 
& $0.090 \scriptscriptstyle \pm \scriptstyle .002$ 
& $0.018 \scriptscriptstyle \pm \scriptstyle .002$ \\
\midrule
MST ($\epsilon=1$) & $\mathbf{0.001\scriptscriptstyle \pm \scriptstyle .001}$ & $0.002\scriptscriptstyle \pm \scriptstyle .001$ & $0.004\scriptscriptstyle \pm \scriptstyle .002$ & $\mathbf{0.001\scriptscriptstyle \pm \scriptstyle .001}$ & $0.002\scriptscriptstyle \pm \scriptstyle .002$ & $0.002\scriptscriptstyle \pm \scriptstyle .001$ \\
PrivSyn ($\epsilon=1$) & $\mathbf{0.001\scriptscriptstyle \pm \scriptstyle .001}$ & $0.002\scriptscriptstyle \pm \scriptstyle .001$ & $0.003\scriptscriptstyle \pm \scriptstyle .001$ & $\mathbf{0.001\scriptscriptstyle \pm \scriptstyle .001}$ & $\mathbf{0.001\scriptscriptstyle \pm \scriptstyle .002}$ & $\mathbf{0.001\scriptscriptstyle \pm \scriptstyle .001}$  \\
PATE-GAN ($\epsilon=1$) &  $0.002\scriptscriptstyle \pm \scriptstyle .001$ & $0.003\scriptscriptstyle \pm \scriptstyle .002$ & $0.002\scriptscriptstyle \pm \scriptstyle .002$ & $\mathbf{0.001\scriptscriptstyle \pm \scriptstyle .001}$ & $\mathbf{0.001\scriptscriptstyle \pm \scriptstyle .002}$ & $\mathbf{0.001\scriptscriptstyle \pm \scriptstyle .001}$  \\
TableDiffusion ($\epsilon=1$) & $0.002\scriptscriptstyle \pm \scriptstyle .001$ & $\mathbf{0.001\scriptscriptstyle \pm \scriptstyle .002}$ & $\mathbf{0.002\scriptscriptstyle \pm \scriptstyle .001}$ & $0.002\scriptscriptstyle \pm \scriptstyle .002$ & $0.002\scriptscriptstyle \pm \scriptstyle .002$ & $\mathbf{0.001\scriptscriptstyle \pm \scriptstyle .001}$ \\
DP-2Stage ($\epsilon=1$) & $\mathbf{0.001\scriptscriptstyle \pm \scriptstyle .001}$ & $0.002\scriptscriptstyle \pm \scriptstyle .002$ & $0.002\scriptscriptstyle \pm \scriptstyle .002$ & $0.002\scriptscriptstyle \pm \scriptstyle .001$ & $0.002\scriptscriptstyle \pm \scriptstyle .001$ & $0.002\scriptscriptstyle \pm \scriptstyle .002$  \\
\midrule
\midrule
\self (lower bound) 
& $0.733 \scriptscriptstyle \pm \scriptstyle .000$ 
& $0.094 \scriptscriptstyle \pm \scriptstyle .000$ 
& $0.125 \scriptscriptstyle \pm \scriptstyle .000$ 
& $0.199 \scriptscriptstyle \pm \scriptstyle .000$ 
& $0.209 \scriptscriptstyle \pm \scriptstyle .000$ 
& $0.273 \scriptscriptstyle \pm \scriptstyle .000$ 
 \\
\bottomrule[0.8pt]
\end{tabular}}

%% file: table/mla.tex
\resizebox{0.75\textwidth}{!}{
\begin{tabular}{lcccccccccccc}
\toprule[0.8pt]
 & \textbf{Adult} & \textbf{Shoppers} & \textbf{Phishing} & \textbf{Magic} & \textbf{Faults} & \textbf{Bean}\\
\midrule
MST & $0.086 \scriptscriptstyle \pm \scriptstyle .001$ & $0.193 \scriptscriptstyle \pm \scriptstyle .002$ & $0.037 \scriptscriptstyle \pm \scriptstyle .003$ & $0.073 \scriptscriptstyle \pm \scriptstyle .001$ & $0.255 \scriptscriptstyle \pm \scriptstyle .002$ & $0.035 \scriptscriptstyle \pm \scriptstyle .003$  \\
PrivSyn & $0.120 \scriptscriptstyle \pm \scriptstyle .003$ & $0.040 \scriptscriptstyle \pm \scriptstyle .001$ & $0.057 \scriptscriptstyle \pm \scriptstyle .002$ & $0.085 \scriptscriptstyle \pm \scriptstyle .003$ & $0.532 \scriptscriptstyle \pm \scriptstyle .001$ & $0.039 \scriptscriptstyle \pm \scriptstyle .002$  \\
TVAE & $0.035 \scriptscriptstyle \pm \scriptstyle .002$ & $0.011 \scriptscriptstyle \pm \scriptstyle .003$ & $0.031 \scriptscriptstyle \pm \scriptstyle .001$ & $0.075 \scriptscriptstyle \pm \scriptstyle .002$ & $0.217 \scriptscriptstyle \pm \scriptstyle .003$ & $0.059 \scriptscriptstyle \pm \scriptstyle .001$ \\
CTGAN & $0.039 \scriptscriptstyle \pm \scriptstyle .003$ & $0.031 \scriptscriptstyle \pm \scriptstyle .002$ & $0.068 \scriptscriptstyle \pm \scriptstyle .001$ & $0.154 \scriptscriptstyle \pm \scriptstyle .003$ & $0.525 \scriptscriptstyle \pm \scriptstyle .002$ & $0.103 \scriptscriptstyle \pm \scriptstyle .001$  \\
TabDDPM & $0.014 \scriptscriptstyle \pm \scriptstyle .001$ & $\mathbf{0.003 \scriptscriptstyle \pm \scriptstyle .002}$ & $0.007 \scriptscriptstyle \pm \scriptstyle .003$ & $0.007 \scriptscriptstyle \pm \scriptstyle .001$ & $\mathbf{0.085 \scriptscriptstyle \pm \scriptstyle .002}$ & $\mathbf{0.003 \scriptscriptstyle \pm \scriptstyle .003}$  \\
TabSyn  
& $0.014 \scriptscriptstyle \pm \scriptstyle .001$ 
& $0.006 \scriptscriptstyle \pm \scriptstyle .002$ 
& $0.025 \scriptscriptstyle \pm \scriptstyle .003$ 
& $\mathbf{0.005 \scriptscriptstyle \pm \scriptstyle .001}$ 
& $0.118 \scriptscriptstyle \pm \scriptstyle .002$ 
& $0.005 \scriptscriptstyle \pm \scriptstyle .001$  \\
GReaT & $0.009 \scriptscriptstyle \pm \scriptstyle .002$ & $0.009 \scriptscriptstyle \pm \scriptstyle .003$ & $0.020 \scriptscriptstyle \pm \scriptstyle .001$ & $0.033 \scriptscriptstyle \pm \scriptstyle .002$ & $0.183 \scriptscriptstyle \pm \scriptstyle .003$ & $0.017 \scriptscriptstyle \pm \scriptstyle .001$  \\
REaLTabFormer 
& $\mathbf{0.004 \scriptscriptstyle \pm \scriptstyle .001}$ 
& $0.004 \scriptscriptstyle \pm \scriptstyle .002$ 
& $\mathbf{0.006 \scriptscriptstyle \pm \scriptstyle .002}$ 
& $0.014 \scriptscriptstyle \pm \scriptstyle .001$ 
& $0.101 \scriptscriptstyle \pm \scriptstyle .003$ 
& $0.006 \scriptscriptstyle \pm \scriptstyle .002$ \\

\midrule
\midrule
MST ($\epsilon=1$) & $\mathbf{0.101 \scriptscriptstyle \pm \scriptstyle .003}$ & $\mathbf{0.048 \scriptscriptstyle \pm \scriptstyle .001}$ & $\mathbf{0.041 \scriptscriptstyle \pm \scriptstyle .002}$ & $\mathbf{0.093 \scriptscriptstyle \pm \scriptstyle .003}$ & $\mathbf{0.489 \scriptscriptstyle \pm \scriptstyle .001}$ & $\mathbf{0.054 \scriptscriptstyle \pm \scriptstyle .002}$  \\
PrivSyn ($\epsilon=1$) & $0.120 \scriptscriptstyle \pm \scriptstyle .002$ & $0.177 \scriptscriptstyle \pm \scriptstyle .003$ & $0.085 \scriptscriptstyle \pm \scriptstyle .001$ & $0.217 \scriptscriptstyle \pm \scriptstyle .002$ & $0.753 \scriptscriptstyle \pm \scriptstyle .003$ & $0.466 \scriptscriptstyle \pm \scriptstyle .001$  \\
PATE-GAN ($\epsilon=1$) & $0.126 \scriptscriptstyle \pm \scriptstyle .001$ & $0.135 \scriptscriptstyle \pm \scriptstyle .002$ & $0.530 \scriptscriptstyle \pm \scriptstyle .003$ & $0.394 \scriptscriptstyle \pm \scriptstyle .001$ & $0.781 \scriptscriptstyle \pm \scriptstyle .002$ & $0.781 \scriptscriptstyle \pm \scriptstyle .003$  \\
TableDiffusion ($\epsilon=1$) & $0.198 \scriptscriptstyle \pm \scriptstyle .002$ & $0.135 \scriptscriptstyle \pm \scriptstyle .003$ & $0.074 \scriptscriptstyle \pm \scriptstyle .001$ & $0.133 \scriptscriptstyle \pm \scriptstyle .002$ & $0.904 \scriptscriptstyle \pm \scriptstyle .003$ & $0.981 \scriptscriptstyle \pm \scriptstyle .001$  \\
DP-2Stage ($\epsilon=1$) & $0.115\scriptscriptstyle \pm \scriptstyle .003$ & $0.072\scriptscriptstyle \pm \scriptstyle .001$ & $0.583\scriptscriptstyle \pm \scriptstyle .004$ & $0.130\scriptscriptstyle \pm \scriptstyle .006$ & $0.721\scriptscriptstyle \pm \scriptstyle .002$ & $0.532\scriptscriptstyle \pm \scriptstyle .003$  \\
\bottomrule[1.0pt]
\end{tabular}}

%% file: table/query.tex
\resizebox{0.75\textwidth}{!}{
\begin{tabular}{lcccccccccccc}
\toprule[0.8pt]
 & \textbf{Adult} & \textbf{Shoppers} & \textbf{Phishing} & \textbf{Magic} & \textbf{Faults} & \textbf{Bean}  \\
\midrule
MST & $0.056\scriptscriptstyle \pm \scriptstyle .018$ & $0.044\scriptscriptstyle \pm \scriptstyle .005$ & $\mathbf{0.009\scriptscriptstyle \pm \scriptstyle .001}$ & $0.035\scriptscriptstyle \pm \scriptstyle .004$ & $0.041\scriptscriptstyle \pm \scriptstyle .003$ & $0.036\scriptscriptstyle \pm \scriptstyle .003$  \\
PrivSyn & $0.009\scriptscriptstyle \pm \scriptstyle .002$ & $0.011\scriptscriptstyle \pm \scriptstyle .006$ & $0.011\scriptscriptstyle \pm \scriptstyle .002$ & $0.011\scriptscriptstyle \pm \scriptstyle .002$ & $0.027\scriptscriptstyle \pm \scriptstyle .004$ & $0.034\scriptscriptstyle \pm \scriptstyle .002$  \\
TVAE & $0.025\scriptscriptstyle \pm \scriptstyle .005$ & $0.034\scriptscriptstyle \pm \scriptstyle .006$ & $0.018\scriptscriptstyle \pm \scriptstyle .000$ & $0.014\scriptscriptstyle \pm \scriptstyle .002$ & $0.026\scriptscriptstyle \pm \scriptstyle .003$ & $0.019\scriptscriptstyle \pm \scriptstyle .001$  \\
CTGAN & $0.015\scriptscriptstyle \pm \scriptstyle .001$ & $0.017\scriptscriptstyle \pm \scriptstyle .001$ & $0.051\scriptscriptstyle \pm \scriptstyle .002$ & $0.037\scriptscriptstyle \pm \scriptstyle .002$ & $0.047\scriptscriptstyle \pm \scriptstyle .006$ & $0.030\scriptscriptstyle \pm \scriptstyle .003$  \\
TabDDPM & $0.006\scriptscriptstyle \pm \scriptstyle .001$ & $0.008\scriptscriptstyle \pm \scriptstyle .001$ & $0.012\scriptscriptstyle \pm \scriptstyle .001$ & $\mathbf{0.006\scriptscriptstyle \pm \scriptstyle .001}$ & $0.021\scriptscriptstyle \pm \scriptstyle .002$ & $\mathbf{0.006\scriptscriptstyle \pm \scriptstyle .001}$ \\
TabSyn  
& $0.005\scriptscriptstyle \pm \scriptstyle .001$ & $0.009\scriptscriptstyle \pm \scriptstyle .001$ & $0.016\scriptscriptstyle \pm \scriptstyle .001$ & $0.007\scriptscriptstyle \pm \scriptstyle .001$ & $\mathbf{0.018\scriptscriptstyle \pm \scriptstyle .003}$ & $0.009\scriptscriptstyle \pm \scriptstyle .001$  \\
GReaT & $0.014\scriptscriptstyle \pm \scriptstyle .002$ & $0.014\scriptscriptstyle \pm \scriptstyle .004$ & $0.049\scriptscriptstyle \pm \scriptstyle .002$ & $0.029\scriptscriptstyle \pm \scriptstyle .003$ & $0.028\scriptscriptstyle \pm \scriptstyle .003$ & $0.011\scriptscriptstyle \pm \scriptstyle .001$ \\
REaLTabFormer & $\mathbf{0.004\scriptscriptstyle \pm \scriptstyle .001}$ & $\mathbf{0.007\scriptscriptstyle \pm \scriptstyle .001}$ & $0.011\scriptscriptstyle \pm \scriptstyle .003$ & $0.012\scriptscriptstyle \pm \scriptstyle .001$ & $0.024\scriptscriptstyle \pm \scriptstyle .002$ & $\mathbf{0.006\scriptscriptstyle \pm \scriptstyle .001}$ \\

\midrule
\midrule
MST ($\epsilon=1$) & $0.071\scriptscriptstyle \pm \scriptstyle .014$ & $0.052\scriptscriptstyle \pm \scriptstyle .017$ & $\mathbf{0.012\scriptscriptstyle \pm \scriptstyle .001}$ & $0.036\scriptscriptstyle \pm \scriptstyle .003$ & $\mathbf{0.045\scriptscriptstyle \pm \scriptstyle .002}$ & $\mathbf{0.037\scriptscriptstyle \pm \scriptstyle .002}$  \\
PrivSyn ($\epsilon=1$) & $\mathbf{0.010\scriptscriptstyle \pm \scriptstyle .001}$ & $0.027\scriptscriptstyle \pm \scriptstyle .007$ & $0.016\scriptscriptstyle \pm \scriptstyle .002$ & $\mathbf{0.025\scriptscriptstyle \pm \scriptstyle .003}$ & $0.100\scriptscriptstyle \pm \scriptstyle .006$ & $0.048\scriptscriptstyle \pm \scriptstyle .004$ \\
PATE-GAN ($\epsilon=1$) & $0.028\scriptscriptstyle \pm \scriptstyle .004$ & $\mathbf{0.024\scriptscriptstyle \pm \scriptstyle .002}$ & $0.117\scriptscriptstyle \pm \scriptstyle .009$ & $0.058\scriptscriptstyle \pm \scriptstyle .005$ & $0.088\scriptscriptstyle \pm \scriptstyle .009$ & $0.191\scriptscriptstyle \pm \scriptstyle .017$ \\
TableDiffusion ($\epsilon=1$) & $0.057\scriptscriptstyle \pm \scriptstyle .006$ & $0.054\scriptscriptstyle \pm \scriptstyle .005$ & $0.071\scriptscriptstyle \pm \scriptstyle .007$ & $0.074\scriptscriptstyle \pm \scriptstyle .011$ & $0.119\scriptscriptstyle \pm \scriptstyle .009$ & $0.052\scriptscriptstyle \pm \scriptstyle .007$ \\
DP-2Stage ($\epsilon=1$) & $0.025\scriptscriptstyle \pm \scriptstyle .003$ & $0.056\scriptscriptstyle \pm \scriptstyle .004$ & $0.085\scriptscriptstyle \pm \scriptstyle .004$ & $0.089\scriptscriptstyle \pm \scriptstyle .003$ & $0.061\scriptscriptstyle \pm \scriptstyle .003$ & $0.055\scriptscriptstyle \pm \scriptstyle .002$  \\
\bottomrule[1.0pt]
\end{tabular}}

%% file: 7.4_exp_ablation.tex
\subsection{In-depth Analysis (RQ3)}\label{sec:exp-indepth}
In this section, we delve into the underlying reasons for the observed performances. Specifically, we employ the proposed fidelity metrics as tools for analyzing the synthesizers' learning process and exploring the impact of privacy budgets on DP synthesizers.

\begin{figure}[t]
    \centering
    \vspace{-5mm}
    \subfigure[CTGAN]
    {
    \includegraphics[width=0.46\linewidth]{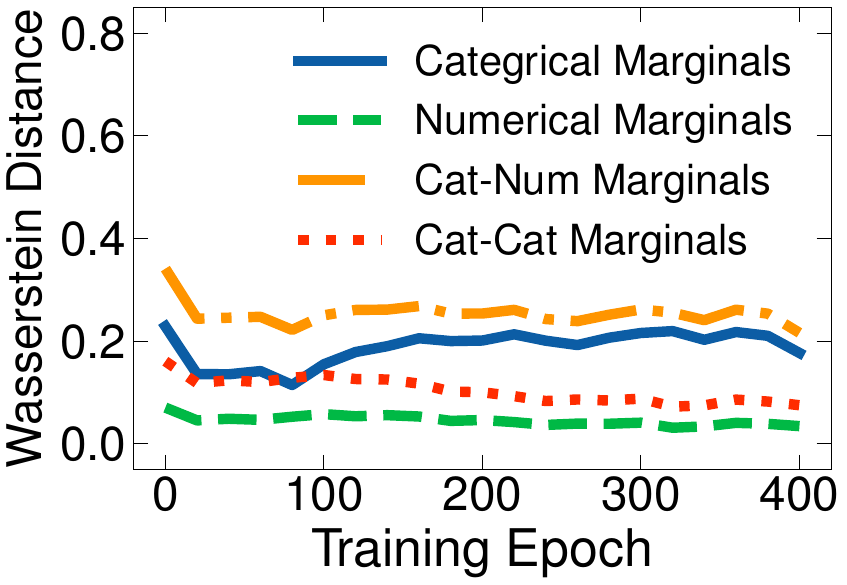}
    \label{fig:audit_ctgan}
    }
    \subfigure[TabDDPM]
    {
    \includegraphics[width=0.46\linewidth]{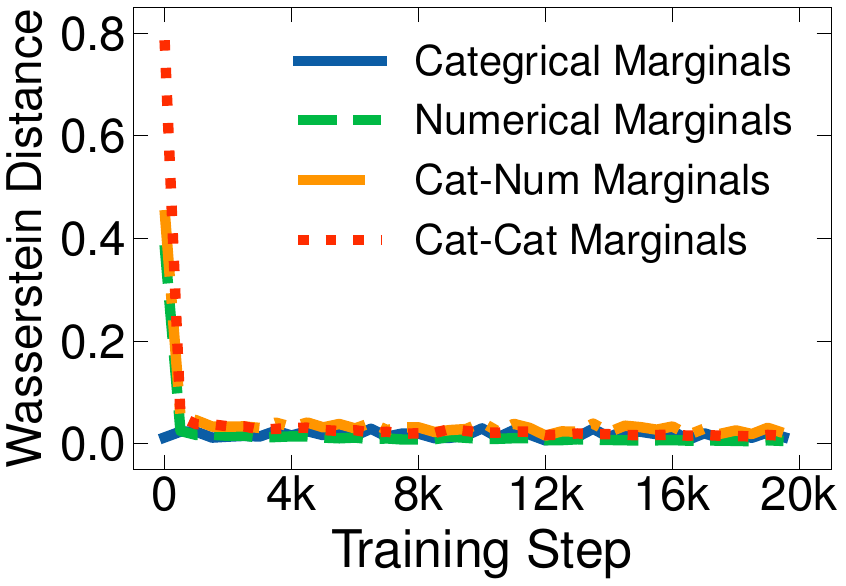}
    \label{fig:audit_tabddpm}
    }
    \vspace{-4mm}
    \caption{\revision{Analyzing the learning process of CTGAN and TabDDPM with proposed fidelity metrics on the Bean dataset.}}
    \label{fig:audit}
\end{figure}

\myquestion{Why Does CTGAN Perform Poorly}
Although CTGAN is widely regarded as a strong synthesizer, our evaluation reveals that it produces the lowest-quality synthetic data. 
This discrepancy raises important questions about the reasons behind CTGAN’s apparent underperformance. 
To investigate this, we scrutinize its learning trajectory, particularly evaluating the fidelity across different marginal types during training. 
As shown in Figure~\ref{fig:audit_ctgan}, both numerical and categorical marginals exhibit unexpected stagnation in improvement.
This suggests that CTGAN’s synthetic data quality is heavily influenced by data preprocessing. Specifically, CTGAN relies on a variational Gaussian mixture model for numerical attributes and conditional sampling for categorical attributes. The model performs well when the data distribution is close to Gaussian; however, most tabular datasets are far more complex and deviate significantly from this assumption~\citep{nips21revisiting}. This mismatch largely explains CTGAN’s suboptimal performance.
Furthermore, this limitation may also account for CTGAN’s strong empirical privacy protections. The model's difficulty in learning complex data structures results in outputs largely independent of any individual training sample, contributing to its good privacy protection.

\begin{figure}
    \centering
    \vspace{-5mm}
   \includegraphics[width=0.46\textwidth]{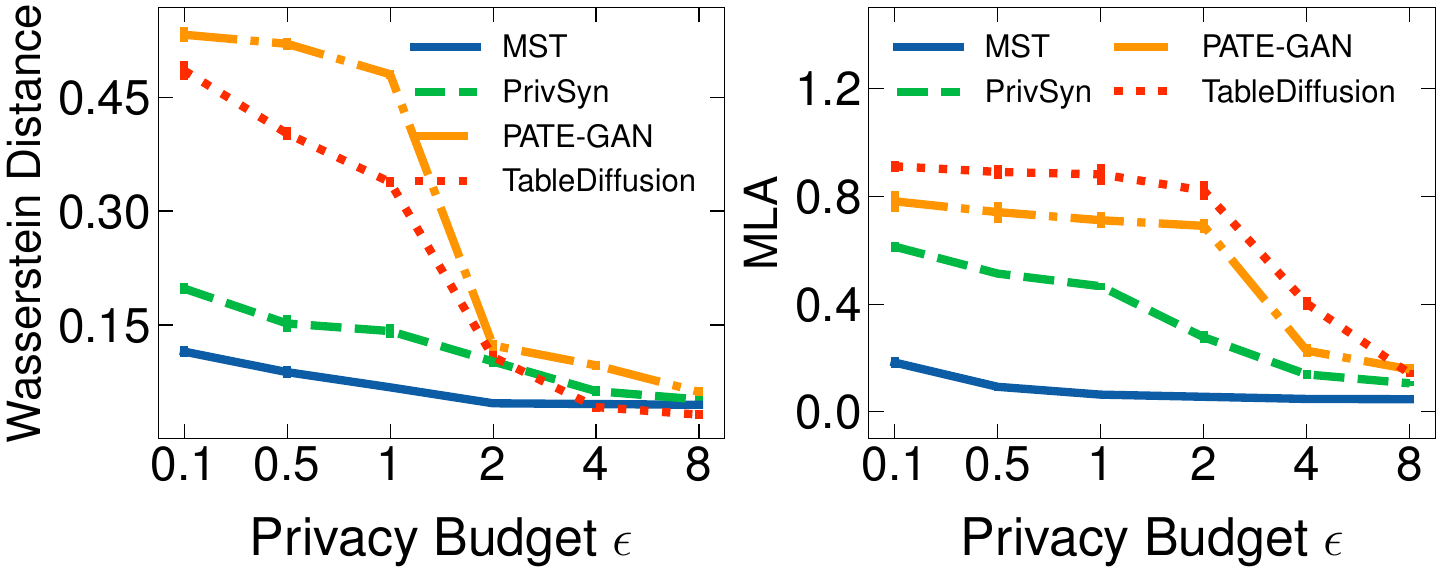}
   \vspace{-4mm}
    \caption{\revision{Impact of privacy budget $\epsilon$ on Bean dataset. The lower score indicates higher fidelity/utility.}}
    \label{fig:budget_2_fig}
\end{figure}


\myquestion{Why Does TabDDPM Excel in HP Synthesizers}
One key finding of our evaluations is the TabDDPM's ability to synthesize high-quality tabular data.
This challenges previous claims that deep generative models generally struggle for tabular data synthesis~\citep {arxiv21benmarkdpsyn}.
We also use proposed fidelity metrics to analyze TabDDPM's learning process. 
As illustrated in Figure~\ref{fig:audit_tabddpm}, the Wasserstein distance across all marginal distributions rapidly decreases, demonstrating the model’s capacity to learn both numerical and categorical distributions.
We attribute this success to the model's architecture: diffusion models have been shown to effectively minimize the Wasserstein distance between synthetic and real data~\citep{nips22diff_wass}. 
This offers a methodological advantage over other generative models, which usually aim to minimize the Kullback-Leibler divergence.
However, despite its strengths, TabDDPM presents significant privacy risks that have been largely overlooked in prior research. Directly applying differential privacy measures would severely degrade the quality of the synthetic data. Nevertheless, diffusion-based methods remain a promising frontier for tabular data synthesis.

\mypara{Large Language Models Are Semantic-aware Synthesizers}
We also notice that the recently emerged LLM-based synthesizer (\ie REaLTabFormer) also shows competitive performance, especially on datasets that consist of rich semantic attributes and complex dependence. 
For instance, REaLTabFormer achieves the best machine learning prediction performance on the Adult dataset, which contains detailed personal information (\eg age and relationship). 
Given the rapid development of LLM and the inherent rich semantics of most tabular data, LLM-based methods may become a new paradigm for realistic data synthesis.


\mypara{The Impact of Privacy Budget}
To analyze the impact of differential privacy on data synthesis, we run DP synthesizers with varying privacy budgets and evaluate their fidelity and utility (see Figure~\ref{fig:budget_2_fig}). 
Our results show that statistical methods (\eg MST) maintain robust performance even with a small privacy budget (\eg $\epsilon=0.5$). 
In contrast, deep generative models typically require much larger privacy budgets (\eg $\epsilon=8$) to achieve comparable results. 
These findings align with previous observations~\citep{arxiv21benmarkdpsyn}, which noted that statistical methods are more resilient to privacy constraints because they rely on estimating a small set of marginals. 

%% file: 0_related.tex
\section{Related Work}
\label{sec:related}

\mypara{Fidelity Evaluation Metrics}
Fidelity is often evaluated based on the distributional similarities of low-order marginals with various statistical measurements. 
Total Variation Distance (TVD)~\citep{iclr24tabsyn} and Kolmogorov-Smirnov Test (KST)~\citep{dankar2022multi} are used to assess univariate distribution similarity for categorical and numerical attributes, respectively. Correlation differences are widely employed for bivariate distributions. Correlation statistics such as Theil's uncertainty coefficient~\citep{acml21ctgabgan,zhao2024ctab}, Pearson correlation~\citep{iclr24tabsyn}, and the correlation ratio~\citep{icml23tabddpm} are utilized to evaluate different types of two-way marginals (categorical, continuous, and mixed).
The main problem with these measures is the lack of versatility. Each type of marginal requires a distinct statistical measure, which complicates the ability to perform a comprehensive comparison across various attribute types.
We refer to Appendix~\ref{appendix:existing_fidelity_metrics} for a detailed discussion of the limitations of existing fidelity metrics.

\mypara{Privacy Evaluation Metrics}
Since HP synthesizers are designed without provable privacy guarantees, privacy evaluation is indispensable for these synthesizers.
Syntactic privacy evaluation metrics (\eg Distance to Closest Records~\citep{acml21ctgabgan}) are the most widely used privacy evaluation for HP synthesizer. 
These metrics compare the input dataset with the output dataset generated by the synthesizer, with closer distances indicating higher privacy risks.
Recently, \citet{arxiv23privcymetric} critiqued these syntactic metrics, highlighting that these ad-hoc metrics can be exploited for reconstruction attacks. 
However, the study did not address the fundamental flaws of these metrics (discussed in Section~\ref{sec:privacy}) and did not introduce new and effective privacy evaluation metrics.
Another way to assess the empirical privacy risks of data synthesis is membership inference attack (MIA)~\citep{sp17membership}. 
Some studies~\citep{usenix22tab_mia,aistats23} have designed different MIA algorithms for tabular data synthesis. However, as shown in Section 5.2, existing MIA algorithms are too weak to differentiate different privacy risks across various synthesizers. Further discussion about existing privacy evaluation metrics can be found in Appendix~\ref{appendix:existing_privacy_metrics}.

\mypara{Utility Evaluation Metrics}
Machine learning prediction and query errors are common downstream tasks for tabular data analysis, and many studies~\citep{usenix21privsyn,nips19ctgan,mckenna21mst} have leveraged these tasks to evaluate the utility of synthetic data. In our evaluation, we also adopt these tasks for utility evaluation and present a reliable metric to address the variability in performance across different machine learning models~\citep{royal2022syn_survey}. Further discussion on utility metrics can be found in Appendix~\ref{appendix:existing_utility_metrics}.

\mypara{Benchmarking Tabular Data Synthesis}
Several studies have benchmarked tabular synthesis algorithms. 
However, they either only focus on DP synthesizers~\citep{arxiv21benmarkdpsyn,sp24ppds}, or neglect the privacy evaluation for HP synthesizers ~\citep{espinosa2023quality,ai22metric,livieris24evaluation,McLachlan18TheAF}.
Additionally, existing benchmarks~\citep{nips24synthcity,arxiv24syntheval} directly leverage existing metrics for evaluation, whereas we identify the limitations of these metrics and propose a new set of evaluation metrics for systematic assessment.

%% file: 8_conclusion.tex
\section{Discussion and Key Takeaways}
\label{sec:takeaway}

Data synthesis has been advocated as an important approach for utilizing data while protecting privacy. Despite the plethora of available data synthesizers, a comprehensive understanding of their strengths and weaknesses remains elusive. 
In this paper, we examine and critique existing metrics, and introduce a systematic framework as well as a new suite of evaluation criteria for assessing data synthesizers. 
We also provide a unified tuning objective to ensure that evaluation results are less affected by accidental choices of synthesizers' hyperparameters.

Our results identify several guidelines for practitioners seeking the best performance for their task and facing the daunting task of selecting and configuring appropriate synthesizers. 

\begin{itemize}
    \item \textit{Model tuning is indispensable.} Tuning hyperparameters can significantly improve synthetic data quality, especially for deep generative models. 
    \item \textit{Statistical methods should be preferred for applications where provable privacy is required.} 
    MST and PrivSyn achieve the best fidelity among DP synthesizers, and they also offer good empirical privacy protection in HP settings.
    \item  \textit{Diffusion models provide the best fidelity and utility.}
    Practitioners are suggested to use diffusion models for tabular synthesis when the quality of synthetic data is the priority over privacy.
    \item \textit{LLMs are semantic-aware synthesizers.} The capability of LLMs enables LLM-based synthesizers to be good at generating tabular data when they consist of rich semantic attributes.
\end{itemize}

Our assessment shows that recently emerged generative models achieve impressive performance on tabular data synthesis and open up new directions in this field. 
At the same time, several critical challenges are also revealed, such as privacy issues of diffusion models and performance gaps between DP and HP synthesizers.
Our evaluation metrics and framework serve a crucial role in highlighting advancements in data synthesis and represent a step toward establishing a standardized evaluation process for this field.



%% file: 9_appendix.tex
\appendix

\input{9.1_app_settings}
\input{9.2_app_results}
\input{9.4_app_tune}

\input{9.3_app_privacy}

\input{9.3_app_fidelity}

\input{9.3_app_utlity}

%% file: 9.1_app_settings.tex
\section{Details of Experimental Setups}

\subsection{Implementation Details}
\label{appendix:implementation}
\mypara{Wasserstein-based Fidelity Metric}
The computation of Wasserstein distance involves solving the linear programming (LR) problem in Equation~\ref{equ:wassertein_opt} and selecting proper marginal distributions.
We compute the Wasserstein distance of all the one-way and two-way marginals and use the mean as the final fidelity score. 
There are many open-source libraries like CVXPY~\citep{16cvxpy} and POT~\citep{21python_pot} that can be used to solve LR reasonably fast. However, when the cost matrix becomes rather large and dense, directly calculating the metric can be computationally expensive. Several options are provided to address this problem: (i) Sinkhorn distance~\citep{nips13sinkhorn} provides a fast approximation to the Wasserstein distance by penalizing the objective with an entropy term. (ii) Sliced-Wasserstein distance~\citep{15sliced}, which uses the Radon transform to linearly project data into one dimension, can be efficiently computed. (iii) Reducing the size of the cost matrix by randomly sampling a small set of points from the probability densities. 
In practice, we find that sampling is both efficient and effective. We randomly sample half of the synthetic data when $n>5,000$ and use the POT library to compute the Wasserstein distance as the fidelity scores.


\mypara{Utility Metrics}
For MLA, we utilize eight machine learning models to compute MLA: SVM, Logistic Regression (or Ridge Regression), Decision Tree, Random Forest, Multilayer Perceptron (MLP), XGBoost~\citep{kdd16xgboost}, CatBoost~\citep{nips18catboost}, and Transformers~\citep{nips21revisiting}. Each model is extensively tuned on real training data to ensure optimal hyperparameters. Performance on classification and regression is evaluated by the F1 score and RMSE, respectively. For query error, we randomly construct 1,000 3-way query conditions and conduct range (point) queries for both synthetic and real data.


\subsection{Datasets}
\label{appendix:dataset}

\begin{itemize}
    \item \textbf{Adult}\footnote{\url{https://archive.ics.uci.edu/dataset/2}} is to predict whether income exceeds 50K/yr.
    \item \textbf{Shoppers}\footnote{\url{https://archive.ics.uci.edu/dataset/468}} is to analyze the intention of online shoppers.
    \item \textbf{Phishing}\footnote{\url{https://archive.ics.uci.edu/dataset/327}} is to predict if a webpage is a phishing site. 
    \item \textbf{Magic}\footnote{\url{https://archive.ics.uci.edu/dataset/159}} is to simulate the registration of high-energy gamma particles in the atmospheric telescope.
    \item \textbf{Faults}\footnote{\url{https://archive.ics.uci.edu/dataset/198}} is the fault detection dataset, which classifies steel plate faults into 7 different types.
    \item \textbf{Bean}\footnote{\url{https://archive.ics.uci.edu/dataset/602}} predicts bean types based on form, shape, and structure. 
    \item \textbf{Obesity}\footnote{\url{https://archive.ics.uci.edu/dataset/544}} estimates the obesity level of individuals.
    \item \textbf{Robot}\footnote{\url{https://archive.ics.uci.edu/dataset/194}} is to predict the robot's move distance around the room.
    \item \textbf{Abalone}\footnote{\url{https://archive.ics.uci.edu/dataset/1}} is to predict the age of abalone.
    \item \textbf{News}\footnote{\url{https://archive.ics.uci.edu/dataset/332}} is to predict the number of shares in social networks.
    \item \textbf{Insurance}\footnote{\url{https://kaggle.com/datasets/tejashvi14/medical-insurance-premium-prediction}} is for prediction on the yearly medical cover cost. 
    \item \textbf{Wine}\footnote{\url{https://archive.ics.uci.edu/dataset/186}} collects physicochemical tests on wine.
\end{itemize}

\input{9.1_hyperparameters_text}

%% file: 9.1_hyperparameters_text.tex
\subsection{Hyperparameter Search Spaces}
\label{appendix:hyperparameters}
\begin{itemize}
    \item \textbf{MST.} Two-way marginals: $\mathrm{Int}[10, 50]$; three-way marginals: $\mathrm{Int}[5, 20]$; bins: $\mathrm{Int}[5, 20]$; max iterations: $\mathrm{Int}[3000, 5000]$.

    \item \textbf{PrivSyn.} Bins: $\mathrm{Int}[5, 20]$; max iterations: $\mathrm{Int}[10, 100]$.

    \item \textbf{TVAE.}  Epochs: $\mathrm{Int}[100, 500]$; batch size: $\mathrm{Int}[500, 5000]$; loss factor: $\mathrm{Float}[1, 5]$; embedding/(de)compression dim: $\mathrm{Int}[128, 512]$; $L_2$ reg: $\mathrm{LogUniform}[1e\text{-}6, 1e\text{-}3]$.

    \item \textbf{CTGAN.} Epochs: $\mathrm{Int}[100, 500]$; batch size: $\mathrm{Int}[500, 5000]$; embedding/generator/discriminator dim: $\mathrm{Int}[128, 512]$; learning rate: $\mathrm{LogUniform}[1e\text{-}5, 1e\text{-}3]$.

    \item \textbf{TabDDPM.} Layers: $\mathrm{Int}[2, 8]$; embedding dim: $\mathrm{Int}[128, 512]$; diffusion timesteps: $\mathrm{Int}[100, 10000]$; training iterations: $\mathrm{Int}[5000, 30000]$; learning rate: $\mathrm{LogUniform}[1e\text{-}5, 3e\text{-}3]$.

    \item \textbf{TabSyn.} Layers: $\mathrm{Int}[1, 4]$; embedding dim: $\mathrm{Int}[128,1024]$.

    \item \textbf{REaLTabFormer.} Epochs: $\mathrm{Int}[100, 1000]$; batch size: $\mathrm{Int}[8, 32]$.

    \item \textbf{GReaT.} Temperature: $\mathrm{Float}[0.6, 0.9]$; fine-tuning epochs: $\mathrm{Int}[100, 300]$; training iterations: $\mathrm{Int}[5000, 30000]$; batch size: $\mathrm{Int}[8, 32]$.

    \item \textbf{PATE-GAN.} Teachers: $\mathrm{Int}[5, 20]$; generator/discriminator layers: $\mathrm{Int}[1, 3]$; generator/discriminator dim: $\mathrm{Int}[50, 200]$; iterations: $\mathrm{Int}[1000, 5000]$; learning rate: $\mathrm{LogUniform}[1e\text{-}5, 1e\text{-}3]$.

    \item \textbf{TableDiffusion.} Layers: $\mathrm{Int}[1, 6]$; diffusion timesteps: $\mathrm{Int}[3, 20]$; epochs: $\mathrm{Int}[5, 20]$; batch size: $\mathrm{Int}[128, 1024]$; noise prediction: \{True, False\}; learning rate: $\mathrm{LogUniform}[1e\text{-}4, 1e\text{-}2]$.

    \item \textbf{DP-2Stage.} fine-tuning epochs: $\mathrm{Int}[1, 5]$.
\end{itemize}


%% file: 9.2_app_results.tex
\begin{table*}[t]
    \centering
    \vspace{-3mm}
    \caption{Fidelity evaluation of synthesizers on training data $D_\text{train}$ of the last six datasets. 
    }
    \vspace{-4mm}
    \label{tab:train_fidelity_2}
    {\include{table/train_fidelity_2}}
\end{table*}

\begin{table*}[t]
    \centering
    \vspace{-3mm}
    \caption{Fidelity evaluation of synthesizers on test data $D_\text{test}$ of the last six datasets.  
    }
    \label{tab:test_fidelity_2}
    \vspace{-4mm}
    {\include{table/test_fidelity_2}}
\end{table*}

\begin{table*}[h]
    \centering
    \vspace{-3mm}
    \caption{\revision{Privacy evaluation of synthesizers on the last six datasets. 
    }}
    \label{tab:privacy_2}
    \vspace{-4mm}
    {\include{table/privacy_2}}
\end{table*}

\begin{table*}[t]
    \centering
    \vspace{-3mm}
    \caption{Utility evaluation (\ie MLA) of synthesizers on the last six datasets. 
    }
    \label{tab:mla_2}
    \vspace{-4mm}
    {\include{table/mla_2}}
\end{table*}

\begin{table*}[h]
    \centering
    \vspace{-3mm}
    \caption{Utility evaluation (\ie query error) of synthesizers on the last six datasets. 
    }
    \label{tab:query_2}
    \vspace{-4mm}
    {\include{table/query_2}}
\end{table*}

\section{Results of Last Six Datasets}

Here we include the results for the remaining six datasets: the fidelity results are in Table~\ref{tab:train_fidelity_2} and Table~\ref{tab:test_fidelity_2}, the privacy results are in Table~\ref{tab:privacy_2}, and the utility results are in Table~\ref{tab:mla_2} and Table~\ref{tab:query_2}.

%% file: table/train_fidelity_2.tex
\resizebox{0.75\textwidth}{!}{
\begin{tabular}{lcccccccccccc}
\toprule[0.8pt]
 & \textbf{Obesity} & \textbf{Robot} & \textbf{Abalone} & \textbf{News} & \textbf{Insurance} & \textbf{Wine} \\
\midrule
MST & $0.041\scriptscriptstyle \pm \scriptstyle .001$ & $0.050\scriptscriptstyle \pm \scriptstyle .002$ & $0.037\scriptscriptstyle \pm \scriptstyle .002$ & $0.060\scriptscriptstyle \pm \scriptstyle .001$ & $0.038\scriptscriptstyle \pm \scriptstyle .005$ & $0.066\scriptscriptstyle \pm \scriptstyle .001$ \\
PrivSyn & $0.034\scriptscriptstyle \pm \scriptstyle .002$ & $0.065\scriptscriptstyle \pm \scriptstyle .012$ & $0.024\scriptscriptstyle \pm \scriptstyle .004$ & $\mathbf{0.018\scriptscriptstyle \pm \scriptstyle .001}$ & $0.033\scriptscriptstyle \pm \scriptstyle .002$ & $0.017\scriptscriptstyle \pm \scriptstyle .000$ \\
TVAE & $0.055\scriptscriptstyle \pm \scriptstyle .004$ & $0.053\scriptscriptstyle \pm \scriptstyle .001$ & $0.048\scriptscriptstyle \pm \scriptstyle .003$ & $0.081\scriptscriptstyle \pm \scriptstyle .001$ & $0.078\scriptscriptstyle \pm \scriptstyle .007$ & $0.039\scriptscriptstyle \pm \scriptstyle .000$ \\
CTGAN & $0.072\scriptscriptstyle \pm \scriptstyle .002$ & $0.106\scriptscriptstyle \pm \scriptstyle .003$ & $0.049\scriptscriptstyle \pm \scriptstyle .004$ & $0.040\scriptscriptstyle \pm \scriptstyle .001$ & $0.090\scriptscriptstyle \pm \scriptstyle .004$ & $0.033\scriptscriptstyle \pm \scriptstyle .001$ \\
TabDDPM & $\mathbf{0.017\scriptscriptstyle \pm \scriptstyle .001}$ & $\mathbf{0.015\scriptscriptstyle \pm \scriptstyle .002}$ & $0.015\scriptscriptstyle \pm \scriptstyle .004$ & $0.034\scriptscriptstyle \pm \scriptstyle .001$ & $0.028\scriptscriptstyle \pm \scriptstyle .005$ & $0.011\scriptscriptstyle \pm \scriptstyle .000$ \\

TabSyn & $0.028\scriptscriptstyle \pm \scriptstyle .003$ & $0.045\scriptscriptstyle \pm \scriptstyle .002$ & $0.020\scriptscriptstyle \pm \scriptstyle .005$ & $0.012\scriptscriptstyle \pm \scriptstyle .002$ & $\mathbf{0.026\scriptscriptstyle \pm \scriptstyle .003}$ & $0.021\scriptscriptstyle \pm \scriptstyle .000$ \\

GReaT$^{1}$ & $0.055\scriptscriptstyle \pm \scriptstyle .005$ & $0.055\scriptscriptstyle \pm \scriptstyle .003$ & $0.022\scriptscriptstyle \pm \scriptstyle .005$ & - & $0.094\scriptscriptstyle \pm \scriptstyle .004$ & $0.019\scriptscriptstyle \pm \scriptstyle .001$ \\
REaLTabFormer & $0.031\scriptscriptstyle \pm \scriptstyle .004$ & $0.029\scriptscriptstyle \pm \scriptstyle .002$ & $\mathbf{0.013\scriptscriptstyle \pm \scriptstyle .004}$ & $0.038\scriptscriptstyle \pm \scriptstyle .001$ & $0.033\scriptscriptstyle \pm \scriptstyle .003$ & $\mathbf{0.008\scriptscriptstyle \pm \scriptstyle .001}$ \\
\midrule
MST ($\epsilon=1$) &  $\mathbf{0.063\scriptscriptstyle \pm \scriptstyle .001}$ & $\mathbf{0.065\scriptscriptstyle \pm \scriptstyle .001}$ & $\mathbf{0.052\scriptscriptstyle \pm \scriptstyle .003}$ & $\mathbf{0.062\scriptscriptstyle \pm \scriptstyle .002}$ & $\mathbf{0.071\scriptscriptstyle \pm \scriptstyle .002}$ & $\mathbf{0.068\scriptscriptstyle \pm \scriptstyle .001}$ \\
PrivSyn ($\epsilon=1$) &  $0.167\scriptscriptstyle \pm \scriptstyle .009$ & $0.169\scriptscriptstyle \pm \scriptstyle .021$ & $0.127\scriptscriptstyle \pm \scriptstyle .009$ & $0.070\scriptscriptstyle \pm \scriptstyle .002$ & $0.124\scriptscriptstyle \pm \scriptstyle .006$ & $0.156\scriptscriptstyle \pm \scriptstyle .004$ \\
PATE-GAN ($\epsilon=1$) &  $0.086\scriptscriptstyle \pm \scriptstyle .003$ & $0.477\scriptscriptstyle \pm \scriptstyle .002$ & $0.331\scriptscriptstyle \pm \scriptstyle .005$ & $0.065\scriptscriptstyle \pm \scriptstyle .002$ & $0.385\scriptscriptstyle \pm \scriptstyle .003$ & $0.251\scriptscriptstyle \pm \scriptstyle .000$ \\
TableDiffusion ($\epsilon=1$) &  $0.347\scriptscriptstyle \pm \scriptstyle .003$ & $0.203\scriptscriptstyle \pm \scriptstyle .001$ & $0.232\scriptscriptstyle \pm \scriptstyle .005$ & $0.135\scriptscriptstyle \pm \scriptstyle .001$ & $0.343\scriptscriptstyle \pm \scriptstyle .002$ & $0.108\scriptscriptstyle \pm \scriptstyle .001$ \\
DP-2Stage ($\epsilon=1$) & $0.075\scriptscriptstyle \pm \scriptstyle .002$ & $0.176\scriptscriptstyle \pm \scriptstyle .002$ & $0.203\scriptscriptstyle \pm \scriptstyle .002$ & $0.163\scriptscriptstyle \pm \scriptstyle .002$ & $0.132\scriptscriptstyle \pm \scriptstyle .002$ & $0.157\scriptscriptstyle \pm \scriptstyle .003$  \\
\midrule
\midrule
\half (upper bound) &  $0.017\scriptscriptstyle \pm \scriptstyle .003$ & $0.010\scriptscriptstyle \pm \scriptstyle .001$ & $0.012\scriptscriptstyle \pm \scriptstyle .004$ & $0.009\scriptscriptstyle \pm \scriptstyle .001$ & $0.026\scriptscriptstyle \pm \scriptstyle .004$ & $0.006\scriptscriptstyle \pm \scriptstyle .000$ \\
\hist (lower bound) &  $0.051\scriptscriptstyle \pm \scriptstyle .001$ & $0.061\scriptscriptstyle \pm \scriptstyle .002$ & $0.069\scriptscriptstyle \pm \scriptstyle .001$ & $0.063\scriptscriptstyle \pm \scriptstyle .002$ & $0.046\scriptscriptstyle \pm \scriptstyle .002$ & $0.068\scriptscriptstyle \pm \scriptstyle .000$ \\
\bottomrule[0.8pt]
\end{tabular}}
\\
\smallskip\scriptsize $^{1}$GReaT cannot be applied to the News dataset because of the maximum length limit of large language models.

%% file: table/test_fidelity_2.tex
\resizebox{0.75\textwidth}{!}{
\begin{tabular}{lcccccccccccc}
\toprule[0.8pt]
 &  \textbf{Obesity} & \textbf{Robot} & \textbf{Abalone} & \textbf{News} & \textbf{Insurance} & \textbf{Wine} \\
\midrule
MST & $0.062\scriptscriptstyle \pm \scriptstyle .003$ & $0.055\scriptscriptstyle \pm \scriptstyle .003$ & $0.062\scriptscriptstyle \pm \scriptstyle .008$ & $0.050\scriptscriptstyle \pm \scriptstyle .004$ & $0.083\scriptscriptstyle \pm \scriptstyle .009$ & $0.075\scriptscriptstyle \pm \scriptstyle .002$ \\
PrivSyn & $0.053\scriptscriptstyle \pm \scriptstyle .005$ & $0.054\scriptscriptstyle \pm \scriptstyle .004$ & $\mathbf{0.032\scriptscriptstyle \pm \scriptstyle .005}$ & $\mathbf{0.018\scriptscriptstyle \pm \scriptstyle .001}$ & $0.074\scriptscriptstyle \pm \scriptstyle .006$ & $0.022\scriptscriptstyle \pm \scriptstyle .001$ \\
TVAE &  $0.059\scriptscriptstyle \pm \scriptstyle .003$ & $0.059\scriptscriptstyle \pm \scriptstyle .007$ & $0.046\scriptscriptstyle \pm \scriptstyle .005$ & $0.079\scriptscriptstyle \pm \scriptstyle .001$ & $0.118\scriptscriptstyle \pm \scriptstyle .009$ & $0.045\scriptscriptstyle \pm \scriptstyle .001$ \\
CTGAN &  $0.085\scriptscriptstyle \pm \scriptstyle .004$ & $0.109\scriptscriptstyle \pm \scriptstyle .009$ & $0.066\scriptscriptstyle \pm \scriptstyle .005$ & $0.040\scriptscriptstyle \pm \scriptstyle .001$ & $0.116\scriptscriptstyle \pm \scriptstyle .008$ & $0.034\scriptscriptstyle \pm \scriptstyle .001$ \\
TabDDPM & $\mathbf{0.043\scriptscriptstyle \pm \scriptstyle .003}$ & $\mathbf{0.028\scriptscriptstyle \pm \scriptstyle .004}$ & $0.034\scriptscriptstyle \pm \scriptstyle .010$ & $0.032\scriptscriptstyle \pm \scriptstyle .001$ & $0.070\scriptscriptstyle \pm \scriptstyle .009$ & $0.017\scriptscriptstyle \pm \scriptstyle .001$ \\
TabSyn & $0.047\scriptscriptstyle \pm \scriptstyle .006$ & $0.050\scriptscriptstyle \pm \scriptstyle .004$ & $0.020\scriptscriptstyle \pm \scriptstyle .006$ & $0.012\scriptscriptstyle \pm \scriptstyle .001$ & $\mathbf{0.067\scriptscriptstyle \pm \scriptstyle .008}$ & $0.028\scriptscriptstyle \pm \scriptstyle .000$ \\
GReaT & $0.062\scriptscriptstyle \pm \scriptstyle .008$ & $0.058\scriptscriptstyle \pm \scriptstyle .006$ & $0.037\scriptscriptstyle \pm \scriptstyle .004$ & - & $0.107\scriptscriptstyle \pm \scriptstyle .010$ & $0.024\scriptscriptstyle \pm \scriptstyle .001$ \\
REaLTabFormer & $0.062\scriptscriptstyle \pm \scriptstyle .006$ & $0.036\scriptscriptstyle \pm \scriptstyle .005$ & $0.040\scriptscriptstyle \pm \scriptstyle .014$ & $0.041\scriptscriptstyle \pm \scriptstyle .001$ & $0.071\scriptscriptstyle \pm \scriptstyle .010$ & $\mathbf{0.015\scriptscriptstyle \pm \scriptstyle .001}$ \\
\midrule
MST ($\epsilon=1$) & $\mathbf{0.075\scriptscriptstyle \pm \scriptstyle .004}$ & $\mathbf{0.072\scriptscriptstyle \pm \scriptstyle .007}$ & $\mathbf{0.080\scriptscriptstyle \pm \scriptstyle .010}$ & $0.051\scriptscriptstyle \pm \scriptstyle .002$ & $\mathbf{0.093\scriptscriptstyle \pm \scriptstyle .006}$ & $\mathbf{0.075\scriptscriptstyle \pm \scriptstyle .001}$ \\
PrivSyn ($\epsilon=1$) & $0.154\scriptscriptstyle \pm \scriptstyle .013$ & $0.177\scriptscriptstyle \pm \scriptstyle .011$ & $0.111\scriptscriptstyle \pm \scriptstyle .011$ & $\mathbf{0.044\scriptscriptstyle \pm \scriptstyle .001}$ & $0.152\scriptscriptstyle \pm \scriptstyle .011$ & $0.130\scriptscriptstyle \pm \scriptstyle .005$ \\
PATE-GAN ($\epsilon=1$) & $0.089\scriptscriptstyle \pm \scriptstyle .004$ & $0.478\scriptscriptstyle \pm \scriptstyle .007$ & $0.353\scriptscriptstyle \pm \scriptstyle .009$ & $0.061\scriptscriptstyle \pm \scriptstyle .002$ & $0.386\scriptscriptstyle \pm \scriptstyle .011$ & $0.250\scriptscriptstyle \pm \scriptstyle .003$ \\
TableDiffusion ($\epsilon=1$) & $0.338\scriptscriptstyle \pm \scriptstyle .005$ & $0.203\scriptscriptstyle \pm \scriptstyle .002$ & $0.226\scriptscriptstyle \pm \scriptstyle .007$ & $0.128\scriptscriptstyle \pm \scriptstyle .001$ & $0.366\scriptscriptstyle \pm \scriptstyle .008$ & $0.098\scriptscriptstyle \pm \scriptstyle .001$ \\
DP-2Stage ($\epsilon=1$) & $0.135\scriptscriptstyle \pm \scriptstyle .002$ & $0.128\scriptscriptstyle \pm \scriptstyle .002$ & $0.147\scriptscriptstyle \pm \scriptstyle .002$ & $0.084\scriptscriptstyle \pm \scriptstyle .003$ & $0.127\scriptscriptstyle \pm \scriptstyle .001$ & $0.157\scriptscriptstyle \pm \scriptstyle .002$  \\
\midrule
\midrule
\half (upper bound) & $0.041\scriptscriptstyle \pm \scriptstyle .006$ & $0.023\scriptscriptstyle \pm \scriptstyle .005$ & $0.028\scriptscriptstyle \pm \scriptstyle .007$ & $0.010\scriptscriptstyle \pm \scriptstyle .002$ & $0.060\scriptscriptstyle \pm \scriptstyle .007$ & $0.014\scriptscriptstyle \pm \scriptstyle .001$ \\
\hist (lower bound) &$0.066\scriptscriptstyle \pm \scriptstyle .002$ & $0.065\scriptscriptstyle \pm \scriptstyle .002$ & $0.094\scriptscriptstyle \pm \scriptstyle .009$ & $0.059\scriptscriptstyle \pm \scriptstyle .006$ & $0.081\scriptscriptstyle \pm \scriptstyle .004$ & $0.076\scriptscriptstyle \pm \scriptstyle .001$ \\
\bottomrule[1.0pt]
\end{tabular}}

%% file: table/privacy_2.tex
\resizebox{0.75\textwidth}{!}{
\begin{tabular}{lcccccccccccc}
\toprule[0.8pt]
 & \textbf{Obesity} & \textbf{Robot} & \textbf{Abalone} & \textbf{News} & \textbf{Insurance} & \textbf{Wine} \\
\midrule
MST 
& $\mathbf{0.013 \scriptscriptstyle \pm \scriptstyle .001}$ 
& $\mathbf{0.008 \scriptscriptstyle \pm \scriptstyle .001}$ 
& $0.030 \scriptscriptstyle \pm \scriptstyle .002$ 
& $0.043 \scriptscriptstyle \pm \scriptstyle .003$ 
& $\mathbf{0.006 \scriptscriptstyle \pm \scriptstyle .001}$ 
& $0.030 \scriptscriptstyle \pm \scriptstyle .002$ \\
PrivSyn 
& $0.027 \scriptscriptstyle \pm \scriptstyle .002$ 
& $0.012 \scriptscriptstyle \pm \scriptstyle .001$ 
& $\mathbf{0.012 \scriptscriptstyle \pm \scriptstyle .003}$ 
& $0.005 \scriptscriptstyle \pm \scriptstyle .002$ 
& $0.013 \scriptscriptstyle \pm \scriptstyle .001$ 
& $\mathbf{0.008 \scriptscriptstyle \pm \scriptstyle .003}$ \\
TVAE 
& $0.104 \scriptscriptstyle \pm \scriptstyle .003$ 
& $0.039 \scriptscriptstyle \pm \scriptstyle .002$ 
& $0.035 \scriptscriptstyle \pm \scriptstyle .001$ 
& $\mathbf{0.004 \scriptscriptstyle \pm \scriptstyle .003}$ 
& $0.036 \scriptscriptstyle \pm \scriptstyle .002$ 
& $0.019 \scriptscriptstyle \pm \scriptstyle .001$ \\
CTGAN  
& $0.026 \scriptscriptstyle \pm \scriptstyle .001$ 
& $0.033 \scriptscriptstyle \pm \scriptstyle .003$ 
& $0.024 \scriptscriptstyle \pm \scriptstyle .002$ 
& $0.007 \scriptscriptstyle \pm \scriptstyle .005$ 
& $0.009 \scriptscriptstyle \pm \scriptstyle .003$ 
& $0.013 \scriptscriptstyle \pm \scriptstyle .001$ \\
TabDDPM 
& $0.333 \scriptscriptstyle \pm \scriptstyle .001$ 
& $0.113 \scriptscriptstyle \pm \scriptstyle .002$ 
& $0.120 \scriptscriptstyle \pm \scriptstyle .003$ 
& $0.008 \scriptscriptstyle \pm \scriptstyle .001$ 
& $0.027 \scriptscriptstyle \pm \scriptstyle .002$ 
& $0.075 \scriptscriptstyle \pm \scriptstyle .003$ \\
TabSyn 
& $0.183 \scriptscriptstyle \pm \scriptstyle .002$ 
& $0.062 \scriptscriptstyle \pm \scriptstyle .001$ 
& $0.102 \scriptscriptstyle \pm \scriptstyle .002$ 
& $0.026 \scriptscriptstyle \pm \scriptstyle .003$ 
& $0.019 \scriptscriptstyle \pm \scriptstyle .002$ 
& $0.124 \scriptscriptstyle \pm \scriptstyle .002$ \\
GReaT 
& $0.263 \scriptscriptstyle \pm \scriptstyle .002$ 
& $0.039 \scriptscriptstyle \pm \scriptstyle .003$ 
& $0.130 \scriptscriptstyle \pm \scriptstyle .001$ 
& - 
& $0.072 \scriptscriptstyle \pm \scriptstyle .002$ 
& $0.034 \scriptscriptstyle \pm \scriptstyle .003$ \\
REaLTabFormer 
& $0.283 \scriptscriptstyle \pm \scriptstyle .002$
& $0.038 \scriptscriptstyle \pm \scriptstyle .001$ 
& $0.150 \scriptscriptstyle \pm \scriptstyle .002$
& $0.008 \scriptscriptstyle \pm \scriptstyle .002$ 
& $0.083 \scriptscriptstyle \pm \scriptstyle .001$
& $0.034 \scriptscriptstyle \pm \scriptstyle .001$
\\
\midrule
MST ($\epsilon=1$) & $0.003\scriptscriptstyle \pm \scriptstyle .001$ & $\mathbf{0.001\scriptscriptstyle \pm \scriptstyle .001}$ & $0.002\scriptscriptstyle \pm \scriptstyle .002$ & $0.002\scriptscriptstyle \pm \scriptstyle .001$ & $0.003\scriptscriptstyle \pm \scriptstyle .001$ & $0.003\scriptscriptstyle \pm \scriptstyle .001$ \\
PrivSyn ($\epsilon=1$) & $0.002\scriptscriptstyle \pm \scriptstyle .001$ & $0.002\scriptscriptstyle \pm \scriptstyle .001$ & $\mathbf{0.001\scriptscriptstyle \pm \scriptstyle .002}$ & $\mathbf{0.001\scriptscriptstyle \pm \scriptstyle .001}$ & $\mathbf{0.002\scriptscriptstyle \pm \scriptstyle .001}$ & $\mathbf{0.001\scriptscriptstyle \pm \scriptstyle .001}$  \\
PATE-GAN ($\epsilon=1$) &  $0.002\scriptscriptstyle \pm \scriptstyle .002$ & $0.003\scriptscriptstyle \pm \scriptstyle .001$ & $0.002\scriptscriptstyle \pm \scriptstyle .001$ & $\mathbf{0.001\scriptscriptstyle \pm \scriptstyle .001}$ & $0.002\scriptscriptstyle \pm \scriptstyle .002$ & $\mathbf{0.001\scriptscriptstyle \pm \scriptstyle .001}$ \\
TableDiffusion ($\epsilon=1$) & $\mathbf{0.002\scriptscriptstyle \pm \scriptstyle .001}$ & $\mathbf{0.001\scriptscriptstyle \pm \scriptstyle .001}$ & $0.002\scriptscriptstyle \pm \scriptstyle .001$ & $0.002\scriptscriptstyle \pm \scriptstyle .001$ & $0.002\scriptscriptstyle \pm \scriptstyle .002$ & $0.003\scriptscriptstyle \pm \scriptstyle .002$ \\
DP-2Stage ($\epsilon=1$) & $0.002\scriptscriptstyle \pm \scriptstyle .002$ & $\mathbf{0.001\scriptscriptstyle \pm \scriptstyle .001}$ & $0.002\scriptscriptstyle \pm \scriptstyle .002$ & $\mathbf{0.001\scriptscriptstyle \pm \scriptstyle .001}$ & $0.002\scriptscriptstyle \pm \scriptstyle .002$ & $0.003\scriptscriptstyle \pm \scriptstyle .002$  \\
\midrule
\midrule
\self (lower bound)  
& $0.671 \scriptscriptstyle \pm \scriptstyle .000$ 
& $0.338 \scriptscriptstyle \pm \scriptstyle .000$ 
& $0.285 \scriptscriptstyle \pm \scriptstyle .000$ 
& $0.068 \scriptscriptstyle \pm \scriptstyle .000$ 
& $0.078 \scriptscriptstyle \pm \scriptstyle .000$ 
& $0.346 \scriptscriptstyle \pm \scriptstyle .000$ \\
\bottomrule[0.8pt]
\end{tabular}}

%% file: table/mla_2.tex
\resizebox{0.75\textwidth}{!}{
\begin{tabular}{lcccccccccccc}
\toprule[0.8pt]
 &  \textbf{Obesity} & \textbf{Robot} & \textbf{Abalone} & \textbf{News} & \textbf{Insurance} & \textbf{Wine} \\
\midrule
MST & $0.332 \scriptscriptstyle \pm \scriptstyle .001$ & $0.146 \scriptscriptstyle \pm \scriptstyle .002$ & $0.096 \scriptscriptstyle \pm \scriptstyle .003$ & $0.498 \scriptscriptstyle \pm \scriptstyle .001$ & $0.270 \scriptscriptstyle \pm \scriptstyle .002$ & $0.347 \scriptscriptstyle \pm \scriptstyle .003$ \\
PrivSyn  & $0.604 \scriptscriptstyle \pm \scriptstyle .003$ & $0.406 \scriptscriptstyle \pm \scriptstyle .001$ & $0.210 \scriptscriptstyle \pm \scriptstyle .002$ & $1.992 \scriptscriptstyle \pm \scriptstyle .003$ & $0.518 \scriptscriptstyle \pm \scriptstyle .001$ & $0.201 \scriptscriptstyle \pm \scriptstyle .002$ \\
TVAE & $0.294 \scriptscriptstyle \pm \scriptstyle .002$ & $0.128 \scriptscriptstyle \pm \scriptstyle .003$ & $0.245 \scriptscriptstyle \pm \scriptstyle .001$ & $0.147 \scriptscriptstyle \pm \scriptstyle .002$ & $0.336 \scriptscriptstyle \pm \scriptstyle .003$ & $0.091 \scriptscriptstyle \pm \scriptstyle .001$ \\
CTGAN & $0.893 \scriptscriptstyle \pm \scriptstyle .003$ & $0.434 \scriptscriptstyle \pm \scriptstyle .002$ & $0.282 \scriptscriptstyle \pm \scriptstyle .001$ & $0.104 \scriptscriptstyle \pm \scriptstyle .003$ & $1.700 \scriptscriptstyle \pm \scriptstyle .002$ & $0.222 \scriptscriptstyle \pm \scriptstyle .001$ \\
TabDDPM & $\mathbf{0.021 \scriptscriptstyle \pm \scriptstyle .001}$ & $\mathbf{0.011 \scriptscriptstyle \pm \scriptstyle .002}$ & $0.043 \scriptscriptstyle \pm \scriptstyle .003$ & $0.047 \scriptscriptstyle \pm \scriptstyle .001$ & $0.140 \scriptscriptstyle \pm \scriptstyle .002$ & $0.047 \scriptscriptstyle \pm \scriptstyle .003$ \\
TabSyn  
& $0.075 \scriptscriptstyle \pm \scriptstyle .001$ 
& $0.086 \scriptscriptstyle \pm \scriptstyle .002$ 
& $\mathbf{0.017 \scriptscriptstyle \pm \scriptstyle .001}$ 
& $\mathbf{0.009 \scriptscriptstyle \pm \scriptstyle .003}$ 
& $\mathbf{0.033 \scriptscriptstyle \pm \scriptstyle .001}$ 
& $0.082 \scriptscriptstyle \pm \scriptstyle .002$ \\
GReaT & $0.117 \scriptscriptstyle \pm \scriptstyle .002$ & $0.050 \scriptscriptstyle \pm \scriptstyle .003$ & $0.038 \scriptscriptstyle \pm \scriptstyle .001$ & - & $0.292 \scriptscriptstyle \pm \scriptstyle .002$ & $0.083 \scriptscriptstyle \pm \scriptstyle .003$ \\
REaLTabFormer 
& $0.054 \scriptscriptstyle \pm \scriptstyle .001$
& $0.017 \scriptscriptstyle \pm \scriptstyle .002$
& $0.020 \scriptscriptstyle \pm \scriptstyle .002$ 
& $0.047 \scriptscriptstyle \pm \scriptstyle .001$
& $0.039 \scriptscriptstyle \pm \scriptstyle .002$ 
& $\mathbf{0.042 \scriptscriptstyle \pm \scriptstyle .003}$ \\
\midrule
\midrule
MST ($\epsilon=1$) & $\mathbf{0.531 \scriptscriptstyle \pm \scriptstyle .003}$ & $\mathbf{0.245 \scriptscriptstyle \pm \scriptstyle .001}$ & $\mathbf{0.241 \scriptscriptstyle \pm \scriptstyle .002}$ & $1.072 \scriptscriptstyle \pm \scriptstyle .003$ & $\mathbf{1.366 \scriptscriptstyle \pm \scriptstyle .001}$ & $0.340 \scriptscriptstyle \pm \scriptstyle .002$ \\
PrivSyn ($\epsilon=1$) & $0.821 \scriptscriptstyle \pm \scriptstyle .002$ & $0.608 \scriptscriptstyle \pm \scriptstyle .003$ & $0.624 \scriptscriptstyle \pm \scriptstyle .001$ & $4.538 \scriptscriptstyle \pm \scriptstyle .002$ & $1.878 \scriptscriptstyle \pm \scriptstyle .003$ & $\mathbf{0.302 \scriptscriptstyle \pm \scriptstyle .001}$ \\
PATE-GAN ($\epsilon=1$) & $0.877 \scriptscriptstyle \pm \scriptstyle .001$ & $0.755 \scriptscriptstyle \pm \scriptstyle .002$ & $2.119 \scriptscriptstyle \pm \scriptstyle .003$ & $\mathbf{0.259 \scriptscriptstyle \pm \scriptstyle .001}$ & $2.325 \scriptscriptstyle \pm \scriptstyle .002$ & $0.405 \scriptscriptstyle \pm \scriptstyle .003$ \\
TableDiffusion ($\epsilon=1$) & $0.968 \scriptscriptstyle \pm \scriptstyle .002$ & $0.439 \scriptscriptstyle \pm \scriptstyle .003$ & $0.287 \scriptscriptstyle \pm \scriptstyle .001$ & $0.781 \scriptscriptstyle \pm \scriptstyle .002$ & $2.503 \scriptscriptstyle \pm \scriptstyle .003$ & $0.489 \scriptscriptstyle \pm \scriptstyle .001$ \\
DP-2Stage ($\epsilon=1$) & $0.721\scriptscriptstyle \pm \scriptstyle .005$ & $0.481\scriptscriptstyle \pm \scriptstyle .003$ & $0.816\scriptscriptstyle \pm \scriptstyle .004$ & $0.751\scriptscriptstyle \pm \scriptstyle .001$ & $1.573\scriptscriptstyle \pm \scriptstyle .003$ & $0.361\scriptscriptstyle \pm \scriptstyle .006$  \\
\bottomrule[1.0pt]
\end{tabular}}

%% file: table/query_2.tex
\resizebox{0.75\textwidth}{!}{
\begin{tabular}{lcccccccccccc}
\toprule[0.8pt]
 & \textbf{Obesity} & \textbf{Robot} & \textbf{Abalone} & \textbf{News} & \textbf{Insurance} & \textbf{Wine} \\
\midrule
MST & $0.035\scriptscriptstyle \pm \scriptstyle .007$ & $0.049\scriptscriptstyle \pm \scriptstyle .005$ & $0.040\scriptscriptstyle \pm \scriptstyle .004$ & $0.033\scriptscriptstyle \pm \scriptstyle .005$ & $0.039\scriptscriptstyle \pm \scriptstyle .004$ & $0.042\scriptscriptstyle \pm \scriptstyle .005$ \\
PrivSyn & $0.027\scriptscriptstyle \pm \scriptstyle .006$ & $0.029\scriptscriptstyle \pm \scriptstyle .003$ & $0.014\scriptscriptstyle \pm \scriptstyle .001$ & $\mathbf{0.010\scriptscriptstyle \pm \scriptstyle .005}$ & $0.035\scriptscriptstyle \pm \scriptstyle .007$ & $0.013\scriptscriptstyle \pm \scriptstyle .002$ \\
TVAE & $0.027\scriptscriptstyle \pm \scriptstyle .003$ & $0.020\scriptscriptstyle \pm \scriptstyle .001$ & $0.016\scriptscriptstyle \pm \scriptstyle .002$ & $0.030\scriptscriptstyle \pm \scriptstyle .006$ & $0.050\scriptscriptstyle \pm \scriptstyle .009$ & $0.028\scriptscriptstyle \pm \scriptstyle .004$ \\
CTGAN & $0.037\scriptscriptstyle \pm \scriptstyle .004$ & $0.033\scriptscriptstyle \pm \scriptstyle .004$ & $0.036\scriptscriptstyle \pm \scriptstyle .005$ & $0.018\scriptscriptstyle \pm \scriptstyle .003$ & $0.055\scriptscriptstyle \pm \scriptstyle .006$ & $0.016\scriptscriptstyle \pm \scriptstyle .003$ \\
TabDDPM & $\mathbf{0.017\scriptscriptstyle \pm \scriptstyle .003}$ & $\mathbf{0.008\scriptscriptstyle \pm \scriptstyle .001}$ & $0.011\scriptscriptstyle \pm \scriptstyle .003$ & $0.017\scriptscriptstyle \pm \scriptstyle .002$ & $\mathbf{0.027\scriptscriptstyle \pm \scriptstyle .007}$ & $0.010\scriptscriptstyle \pm \scriptstyle .001$ \\
TabSyn   & $0.020\scriptscriptstyle \pm \scriptstyle .002$ & $0.017\scriptscriptstyle \pm \scriptstyle .003$ & $\mathbf{0.009\scriptscriptstyle \pm \scriptstyle .002}$ & $0.005\scriptscriptstyle \pm \scriptstyle .001$ & $0.027\scriptscriptstyle \pm \scriptstyle .006$ & $0.016\scriptscriptstyle \pm \scriptstyle .001$ \\
GReaT & $0.030\scriptscriptstyle \pm \scriptstyle .003$ & $0.014\scriptscriptstyle \pm \scriptstyle .001$ & $0.019\scriptscriptstyle \pm \scriptstyle .003$ & - & $0.041\scriptscriptstyle \pm \scriptstyle .007$ & $0.013\scriptscriptstyle \pm \scriptstyle .001$ \\
REaLTabFormer & $0.027\scriptscriptstyle \pm \scriptstyle .004$ & $0.009\scriptscriptstyle \pm \scriptstyle .001$ & $0.015\scriptscriptstyle \pm \scriptstyle .002$ & $0.019\scriptscriptstyle \pm \scriptstyle .002$ & $0.032\scriptscriptstyle \pm \scriptstyle .006$ & $\mathbf{0.007\scriptscriptstyle \pm \scriptstyle .001}$ \\

\midrule
\midrule
MST ($\epsilon=1$) & $0.043\scriptscriptstyle \pm \scriptstyle .005$ & $\mathbf{0.050\scriptscriptstyle \pm \scriptstyle .008}$ & $\mathbf{0.041\scriptscriptstyle \pm \scriptstyle .003}$ & $0.043\scriptscriptstyle \pm \scriptstyle .004$ & $\mathbf{0.033\scriptscriptstyle \pm \scriptstyle .004}$ & $\mathbf{0.045\scriptscriptstyle \pm \scriptstyle .004}$ \\
PrivSyn ($\epsilon=1$) & $0.060\scriptscriptstyle \pm \scriptstyle .004$ & $0.095\scriptscriptstyle \pm \scriptstyle .002$ & $0.051\scriptscriptstyle \pm \scriptstyle .003$ & $\mathbf{0.027\scriptscriptstyle \pm \scriptstyle .005}$ & $0.062\scriptscriptstyle \pm \scriptstyle .007$ & $0.064\scriptscriptstyle \pm \scriptstyle .008$ \\
PATE-GAN ($\epsilon=1$) & $\mathbf{0.037\scriptscriptstyle \pm \scriptstyle .001}$ & $0.150\scriptscriptstyle \pm \scriptstyle .023$ & $0.223\scriptscriptstyle \pm \scriptstyle .032$ & $0.029\scriptscriptstyle \pm \scriptstyle .003$ & $0.138\scriptscriptstyle \pm \scriptstyle .011$ & $0.158\scriptscriptstyle \pm \scriptstyle .013$ \\
TableDiffusion ($\epsilon=1$) & $0.108\scriptscriptstyle \pm \scriptstyle .010$ & $0.071\scriptscriptstyle \pm \scriptstyle .012$ & $0.085\scriptscriptstyle \pm \scriptstyle .010$ & $0.050\scriptscriptstyle \pm \scriptstyle .003$ & $0.195\scriptscriptstyle \pm \scriptstyle .011$ & $0.048\scriptscriptstyle \pm \scriptstyle .006$ \\
DP-2Stage ($\epsilon=1$) & $0.518\scriptscriptstyle \pm \scriptstyle .003$ & $0.572\scriptscriptstyle \pm \scriptstyle .002$ & $0.498\scriptscriptstyle \pm \scriptstyle .003$ & $0.029\scriptscriptstyle \pm \scriptstyle .002$ & $1.071\scriptscriptstyle \pm \scriptstyle .002$ & $0.052\scriptscriptstyle \pm \scriptstyle .003$  \\
\bottomrule[1.0pt]
\end{tabular}}

%% file: 9.4_app_tune.tex
\section{Impact of Model Tuning Phase}
\label{appendix:model_tuning}

\mypara{\revision{Incorpating MDS for model tuning}}
\revision{The tuning objective in Equation~\ref{equ:tune_obj} does not incorporate the proposed privacy metric. In Table~\ref{tab:tune_w_mds}, we present the relative performance differences when incorporating MDS into model tuning versus not incorporating it. The results indicate that while privacy evaluation performance shows slight improvement, the overall impact of including MDS in the tuning process is negligible.}

\mypara{Comparison with Existing Tuning Objective}
We conduct experiments to show the effectiveness of our tuning objective. 
Specifically, we consider the following objectives and metrics:

\begin{itemize}
    \item Existing Tuning Objectives. 
    We adopt the tuning method of TabDDPM, which uses the machine learning efficiency of synthetic data on CatBoost as its tuning objective (we call it MLE$_\text{obj}$).    
    \item Existing Evaluation Metrics. We evaluated the results using five widely used fidelity metrics, including Total Variation Distance (TVD) and Kolmogorov-Smirnov Test (KST), Theil's uncertainty coefficient, Pearson correlation, and the correlation ratio. For existing utility metrics, we included machine learning efficiency on CatBoost and query errors. 
\end{itemize}

Table~\ref{tab:tune_effectiveness} reports the performance improvements of the existing tuning objective (\ie MLE$_{\text{obj}}$) and the proposed method across various evaluation metrics on TabDDPM.
The results indicate that our proposed tuning objective significantly enhances performance on both the proposed and existing metrics. 
Additionally, while MLE$_{\text{obj}}$ effectively improves machine learning efficiency (which is also their optimization objective), it shows limited improvement in other aspects, such as all the fidelity metrics and query errors.

\mypara{Impact of Different Coefficient Configurations}
We also present the results of various coefficient combinations during the tuning phase, as shown in Table~\ref{tab:tune_coeffecient}. The results demonstrate that our tuning objective is highly robust to different coefficient assignments, with all combinations showing a significant improvement over the default settings. Additionally, we note that practitioners can adjust these coefficients based on specific application needs to enhance certain characteristics of the synthetic data. For example, one may want to increase $\alpha_2$ to improve the quality of synthetic data for model selection tasks. We also observed that no single coefficient configuration maximizes model performance across all metrics.

\begin{table}[t]
    \centering
    \vspace{-5mm}
    \caption{Performance improvements (\%) for TabDDPM when training with the proposed tuning objective.}
    \vspace{-4mm}
    {\small \include{table/tune_coeffecient}}
    \label{tab:tune_coeffecient}
\end{table}

\begin{table}[t]
    \centering
    \vspace{-5mm}
    \caption{\revision{Relative performance improvements (\%) when tuning with the proposed privacy metric (\ie MDS).}}
    \vspace{-4mm}
    {\small \include{table/tune_w_mds}}
    \label{tab:tune_w_mds}
\end{table}


%% file: table/tune_coeffecient.tex
\resizebox{0.4\textwidth}{!}{\begin{tabular}{ccc|ccccc}
    \toprule
    \multicolumn{1}{c}{\multirow{2}{*}{$\mathbf{\alpha}_1$}} & \multicolumn{1}{c}{\multirow{2}{*}{$\mathbf{\alpha}_2$}} & \multicolumn{1}{c|}{\multirow{2}{*}{$\mathbf{\alpha}_3$}} & \multicolumn{2}{c}{\textbf{Fidelity} $\uparrow$} & \multicolumn{2}{c}{\textbf{Utility $\uparrow$}} \\
    \cmidrule(lr){4-5}  \cmidrule(lr){6-7}
    & & & $D_\text{Train}$ & $D_\text{Test}$ & MLA & Query Error \\
    \midrule
    0 & 1/2 & 1/2 & $10.57$ & $10.01$ & $8.45$ & $7.90$ \\
    1/4 & 1/2 & 1/4 & $11.17$ & $10.48$ & $8.30$ & $7.21$ \\
    1/4 & 1/4 & 1/2 & $11.24$ & $10.33$ & $8.08$ & $7.91$ \\
    1/3 & 1/3 & 1/3 & $11.34$ & $10.95$ & $8.32$ & $7.86$ \\
    1/2 & 1/4 & 1/4 & $12.16$ & $10.98$ & $7.64$ & $7.06$ \\
    1/2 & 0 & 1/2 & $11.34$ & $10.23$ & $7.15$ & $7.65$ \\
    1/2 & 1/2 & 0 & $10.38$ & $9.97$ & $8.62$ & $7.17$ \\
    \bottomrule
\end{tabular}}

%% file: table/tune_w_mds.tex
\resizebox{0.46\textwidth}{!}{\begin{tabular}{cccccc}
    \toprule
    \multicolumn{1}{c}{\multirow{2}{*}{\textbf{Synthesizer}}} & \multicolumn{2}{c}{\textbf{Fidelity} $\uparrow$} & \multicolumn{1}{c}{\textbf{Privacy} $\uparrow$} & \multicolumn{2}{c}{\textbf{Utility $\uparrow$}} \\
    \cmidrule(lr){2-3} \cmidrule(lr){4-4} \cmidrule(lr){5-6}
    & $D_\text{Train}$ & $D_\text{Test}$ & MDS & MLA & Query Error \\
    \midrule
    MST & $0.05\%$ & $0.02\%$ & $0.03\%$ & $0.02\%$ & $0.04\%$ \\
    PrivSyn & $-0.02\%$ & $-0.04\%$ & $0.02\%$ & $-0.01\%$ & $-0.03\%$ \\
    TVAE & $-0.03\%$ & $-0.05\%$ & $0.04\%$ & $-0.05\%$ & $-0.02\%$ \\
    CTGAN & $0.01\%$ & $0.02\%$ & $0.02\%$ & $0.01\%$ & $0.03\%$ \\
    PATE-GAN & $-0.02\%$ & $-0.02\%$ & $0.03\%$ & $0.02\%$ & $-0.01\%$ \\
    TabDDPM & $0.03\%$ & $0.05\%$ & $0.06\%$ & $0.04\%$ & $0.03\%$ \\
    TableDiffusion & $-0.03\%$ & $-0.05\%$ & $0.02\%$ & $-0.06\%$ & $-0.05\%$ \\
    REaLTabFormer & $-0.08\%$ & $-0.06\%$ & $0.01\%$ & $-0.04\%$ & $-0.03\%$ \\
    \bottomrule
\end{tabular}}

%% file: 9.3_app_privacy.tex
\section{Comparing MDS with Other Privacy Metrics}

\subsection{Existing Privacy Metrics and Limitations}
\label{appendix:existing_privacy_metrics}

\mypara{Syntactic Privacy Evaluation Metrics}
Researchers propose to measure the privacy risk of synthetic data by comparing an input dataset with the output dataset generated by the synthesizer, typically using the distances between data records.  For example, the Distance to Closest Records (DCR)~\citep{acml21ctgabgan} metric looks at the distribution of the distances from each synthetic data point to its nearest real one, and uses the 5th percentile (or the mean) of this distribution as the privacy score.   
A small score is interpreted as indicating that the synthetic dataset is too close to real data, signaling a high risk of information leakage.  
DCR and/or other similar metrics are widely used both in academia~\citep{yale19assessing} and industry~\citep{aws22,gretel23}, and have become the conventional privacy metrics for HP synthesizers~\citep{royal2022syn_survey}. 

Metrics such as DCR are \textbf{syntactic} since they are computed based on the datasets instead of the synthesis algorithm.
We note that when researchers were studying privacy properties of data anonymizers, syntactic privacy metrics such as $k$-anonymity~\citep{k-anonymity}, $\ell$-diversity~\citep{l-diversity}, and $t$-closeness~\citep{icde06tcloseness} were introduced.  Similarly, these metrics consider only the anonymized dataset (and not the algorithm generating the dataset) when measuring privacy.  
Over the last decade and a half, the community has gradually recognized the limitations of such syntactic privacy evaluation metrics and adopted privacy notions such as differential privacy~\citep{dwork_dp}, which defines privacy as a property of the data processing algorithm, instead of the property of a particular output. 

\mypara{Limitations of DCR}
We use the DCR as an example to show the limitations of such syntactic metrics.
First, DCR \textit{overestimates} the privacy risks when data points are naturally clustered close together.  
As illustrated by discussions about DP~\citep{dwork14book,ninghui17book}, leaking information regarding an individual should not be considered a privacy violation if the leakage can occur even if the individual's data is not used.  
Analogously, having some synthetic data very close to real ones does \textit{not} mean worse privacy if this situation can occur even if each data point is removed~\citep{dwork14book,ninghui17book}.  
Consider, for example, a dataset that is a mixture of two Gaussians with small standard deviations. A good synthetic dataset is likely to follow the same distribution and has many data points close to the real ones.  
DCR interprets this closeness as a high privacy risk, overlooking the fact that the influence of any individual instance on synthetic data is insignificant.

Second, DCR measures privacy loss using the 5th percentile (or mean) proximity to real data, which fails to bound the \textit{worst-case} privacy leakage among all records. 
When measuring the privacy leakage across different individuals, one needs to ensure that the worst-case leakage is bounded, so that every individual's privacy is protected.  It is unacceptable to use a mechanism that sacrifices the privacy of some individuals, even though the protection averaged over the population is good.  This point is illustrated by the fact that the re-identification of one or a few individuals is commonly accepted as privacy breaches~\citep{ccs13membership}.  

\mypara{Membership Inference Attack on Data Synthesis}
MIA has been widely used as an empirical privacy evaluation in machine learning, which has been extensively studied on discriminative models~\citep{sp17membership,sp22lira}. 
For generative models like diffusion models~\citep{icml23diff_mia} and LLM~\citep{duan24membership}, studies mainly focus on both the \textit{white-box} setting (where an adversary has full access to the trained model) and on the \textit{black-box} setting (where an adversary is aware of the type of generative model). In the realm of data synthesis, \citet{usenix24recon} claims that the \textit{non-box} setting should be considered in practice: the adversary has access to the synthetic dataset but no information about the underlying generative model or even the specifications of the synthetic data generation algorithm. \citet{usenix22tab_mia} performs the first non-box membership inference attack called Groundhog, which utilizes handcrafted features extracted from synthetic data distribution to train shadow models.  
TAPAS~\citep{houssiau2022tapas} utilizes target counting queries as features and trains a random forest classifier to perform the attack and achieve better performance than~\citet{usenix22tab_mia}.
DOMIAS~\citep{aistats23} utilizes the additional reference dataset to calibrate the density estimation of output distributions and achieve state-of-the-art performance for data synthesis. 
However, our experiments show that TAPAS and DOMIAS are still insufficient to distinguish nuances of privacy risks in all scenarios.

\subsection{Discussion of MDS}
\label{appendix:diss_mds}

\mypara{Comparison with Syntactic Privacy Evaluation Metrics} 
We use DCR as an example to show how the proposed membership disclosure score addresses the drawbacks of syntactic metrics. First, MDS addresses DCR's over-estimating leakage issue by quantifying how much, including each record $x$ \textit{changes} the distance between $x$ and the closest synthetic data. If including $x$ results in records much closer to $x$ being generated, then the disclosure risk is high.  Conversely, if records close to $x$ are generated regardless of whether $x$ is included, then the disclosure risk for $x$ is low. Therefore, MDS follows a distinguishing game designed to mirror the DP definition, rather than relying on the density of data points. Additionally, MDS uses the maximum disclosure risk among all records, providing a stable \textit{worst-case} privacy measurement.

\mypara{Connections to MIAs}
Both MDS and MIAs measure privacy risks by assessing the influence of discrepancies observed in the synthesizer when trained with or without certain records. 
Additionally, MDS incorporates shadow model techniques~\citep{sp17membership} to estimate the influence for all data records, which is the standard approach in MIAs. 
However, unlike MIAs, MDS directly assesses the privacy risks of training data without relying on the construction of the membership inference security game~\citep{sp22lira}. 
Consequently, MDS's privacy estimation does not depend on the effectiveness of one specific attack algorithm, offering greater flexibility in evaluating various types of data synthesizers.



\mypara{Limitations of MDS}
Although MDS provides a straightforward way to assess the privacy risk of data synthesis, it may not apply to all synthesizers. 
Pathological synthesizers exist for which MDS is inappropriate.  
For example, a synthesizer maps all data points $x\in D$ to their opposites: $x \mapsto-x$. 
Suppose the nearest neighbor to $x$ in the real dataset is $x+\varepsilon$. 
In this case, MDS would be proportional to $|d(x, s(x))-d(x, s(x+\varepsilon))|$, which can be tricked arbitrarily small with $\varepsilon$.
However, this synthesizer completely reveals the dataset and MDS would suggest a false sense of privacy. 
Therefore, while we find MDS to be effective in assessing privacy risks for the synthesizers we tested, caution should be exercised when applying it in practice. For scenarios where privacy is paramount, we highly recommend using DP synthesizers instead of HP synthesizers.



\subsection{Experimental Comparison of Privacy Metrics}
\label{appendix:comparsion_privacy_metrics}

\begin{table*}[t]
    \centering
    \vspace{-3mm}
    \caption{Comparison of privacy evaluation metrics for HP synthesizers on Adult. $D_t$, $D_r$, and $S$ are the training, reference, and synthetic data. $\mathsf{A}$ is the synthesis algorithm. Syntactic metrics (DCR and NNDR) are too unstable to provide meaningful privacy measures. MIAs (Groundhog, TAPAS, and MODIAS) fail to distinguish the different levels of privacy risks of synthesizers.}
    \vspace{-4mm}
    {\small \include{table/metric_compare}}
    \label{tab:privacy_metric_compare}
\end{table*}

\mypara{Comparison with Syntactic Metrics and MIAs}
We compare the efficacy of different privacy metrics by conducting proof-of-concept experiments. Specifically, we consider the following metrics:
\begin{itemize}
    \item \textbf{DCR}~\citep{acml21ctgabgan} measures the distance between the synthetic record and its closest real neighbor. The 5th percentile of the distance distribution represents the privacy score. We also utilize the worst-case (nearest distance) of DCR for comparison. 
    \item \textbf{NNDR}~\citep{acml21ctgabgan} calculates the distance ratio between the closest and second closest real neighbor to synthetic data. 
    \item \textbf{Groundhog}~\citep{usenix22tab_mia} calculates statistics from synthetic data as features and uses these features to train shadow models for attack.
    \item \textbf{TAPAS}~\citep{houssiau2022tapas} leverages the counting queries as features and trains a random forest classifier for membership inference.
    \item \textbf{DOMIAS}~\citep{aistats23} utilizes a reference dataset to perform the attack via a likelihood ratio test.
\end{itemize}

We randomly divide the dataset into two disjoint subsets: a training set $D_t$ and a reference set $D_r$, where $|D_t| = |D_r|$ and they share the same data distribution. 
Each synthesizer is trained on $D_t$ and generates the synthetic data $D$, while the reference data $D_r$ remains unused during the synthesis process.
Different treatments are applied for different metrics: 
(i) For syntactic metrics (\ie DCR and NNDR), we compute the privacy score by treating either $D_t$ or $D_r$ as the real data. Unless a synthesizer provides very good privacy, it is expected that the privacy leakage on $D_r$ is significantly smaller than that on $D_t$, since the synthetic data is generated using $D_t$ and is independent of $D_r$. 
(ii) For MIAs and MDS, training dataset $D_t$ and synthesis algorithm $\mathsf{A}$ are utilized to compute the privacy leakage. Table~\ref{tab:privacy_metric_compare} presents the results for HP synthesizers on Adult. We have the following observations: 
\begin{itemize}
    \item \textit{Syntactic metrics are not stable.} The standard deviations of syntactic metrics are quite large compared to their mean values. This instability is pronounced when using the nearest distance as the score, representing the worst-case assessment. 
    \item \textit{Syntactic metrics are improper privacy measurements.} 
    When using training data $D_t$ or reference data $D_r$ as real data to compute DCR and NNDR, the score differences are very small compared to their standard deviations.  We note that for a good privacy evaluation metric, only when a synthesizer provides a very strong privacy guarantee would we expect the two scores to be very similar. 
    Since it is impossible that all HP synthesizers can provide such a high level of strong privacy guarantee, we assert this is because these syntactic metrics do not provide a good measure of privacy. 
    \item \textit{MIAs fail to distinguish different levels of privacy.} 
    Experimental results show that the performance of MI attacks is relatively low for most synthesizers. We attribute the failure to the inherent randomness of synthesizers and synthetic datasets, which make it difficult to capture reliable signals to determine the membership. 
    \item \textit{MDS is a reliable privacy evaluation metric.} It is observed that the variance of MDS is very small, indicating its robustness for assessing data synthesizers. Additionally, MDS can also detect subtle differences in privacy leakage across various HP synthesizers.
\end{itemize}

\begin{figure}[th]
    \centering
    \vspace{-5mm}
    \includegraphics[width=0.45\linewidth]{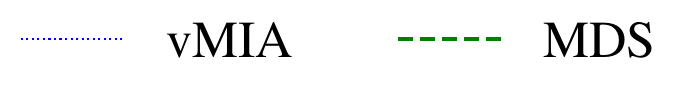}
    \vspace{-2mm}
    \\
    \subfigure[MST]
    {
    \centering
    \includegraphics[width=0.45\linewidth]{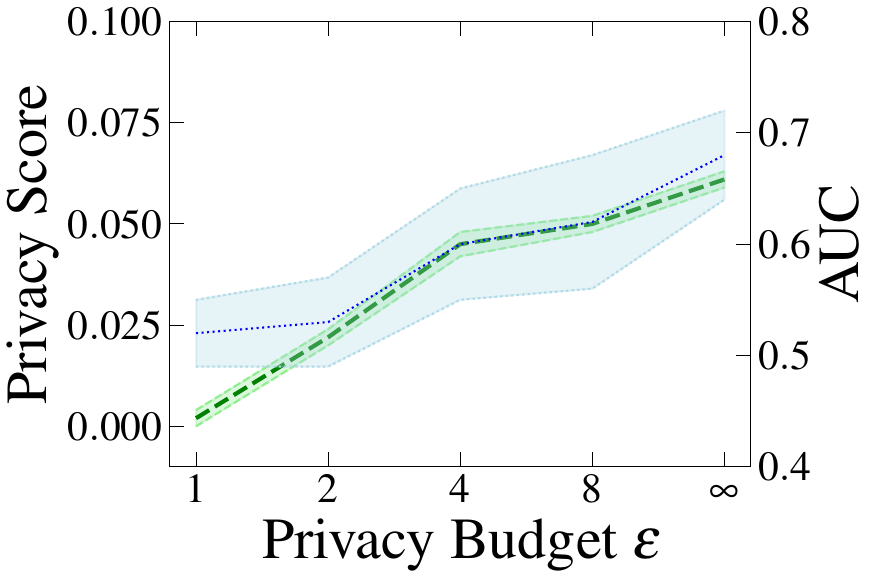}
    \label{fig:mds_mst}
    }
    \subfigure[PATE-GAN]
    {
    \centering
    \includegraphics[width=0.45\linewidth]{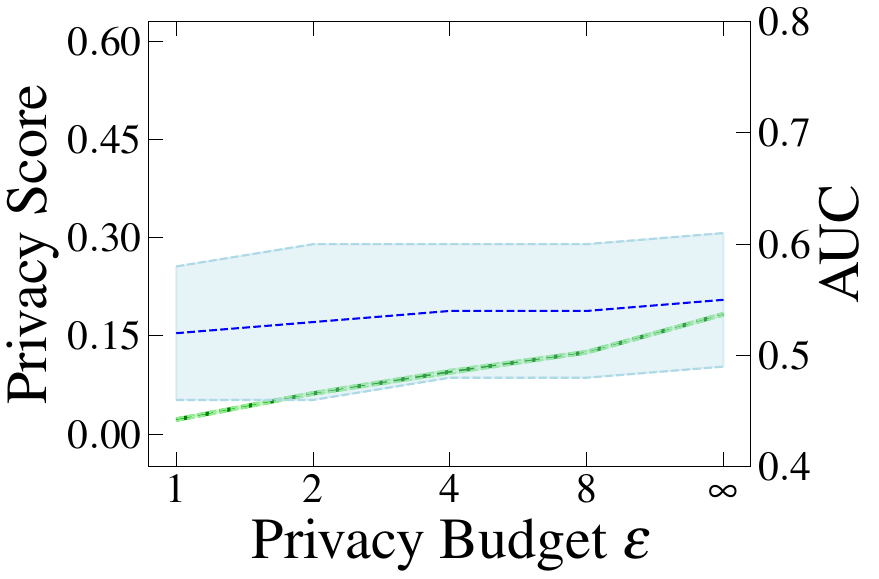}
    \label{fig:mds_pategan}
    }
    \vspace{-4mm}
    \caption{Privacy evaluation comparison between vMIA~\citep{meeus2023achilles} and MDS. MDS uses the left y-axis (``Privacy Score'') whereas vMIA uses right y-axis (``AUC'').}
\label{fig:mds_meeus}
\end{figure}

\begin{table*}[h]
    \centering
    \vspace{-3mm}
    \caption{Performance improvements (\%) of existing tuning objective (\ie MLE$_{\text{obj}}$) and the proposed one on TabDDPM.}
    \vspace{-4mm}
        \begin{threeparttable}
    {\include{table/tune_effectiveness}}
    \end{threeparttable}
    \label{tab:tune_effectiveness}
\end{table*}

\mypara{Comparison with Meeus et al}
\citet{meeus2023achilles} proposed a new approach to evaluate the empirical privacy risks. 
It first identifies vulnerable samples by examining their closeness and then conducts a shadow model-based membership inference attack on the vulnerable sample for evaluation. 
While this approach (we call it vMIA) does not align with the standard MIA setting, it may serve as a viable tool for privacy evaluation. 
Thus, we conduct the following experiments to compare its effectiveness.
Specifically, we train two DP synthesizers (\ie MST and PATE-GAN) with varying levels of privacy protection, and measure the empirical privacy risk using vMIA and MDS. 
The results of both experiments are presented in Figure~\ref{fig:mds_meeus}. We observe an improvement in attack performance for MST, whereas the performance of PATE-GAN remains relatively low (below 60\% AUC) across all levels of privacy budgets.
Furthermore, vMIA suffers from relatively high variance.
In contrast, MDS reliably detects different privacy risks across both evaluated models.

%% file: table/metric_compare.tex
\resizebox{0.9\textwidth}{!}{
\begin{tabular}{ccccccccc}
\toprule[0.8pt]
\textbf{Privacy Evaluation Metric} & \textbf{Metric Input} & \textbf{MST} $(\epsilon=\infty)$ & \textbf{PrivSyn} $(\epsilon=\infty)$ & \textbf{TVAE} & \textbf{CTGAN}& \textbf{TabDDPM} & \textbf{GReaT}  \\
\midrule

\multirow{2}{*}{\begin{tabular}[c]{@{}c@{}} DCR \\ (5th percentile distance) \end{tabular}} 
& $D_t, S$
& $0.535 \scriptscriptstyle \pm \scriptstyle .121$ &  $0.520 \scriptscriptstyle \pm \scriptstyle .182$ & $0.493 \scriptscriptstyle \pm \scriptstyle .116$ & $0.533 \scriptscriptstyle \pm \scriptstyle .103$ & $0.409 \scriptscriptstyle \pm \scriptstyle .181$ & $0.437 \scriptscriptstyle \pm \scriptstyle .122$  \\ 
\cmidrule(lr){2-8} 
& $D_r, S$ & $0.527 \scriptscriptstyle \pm \scriptstyle .146$ &  $0.531 \scriptscriptstyle \pm \scriptstyle .194$ & $0.487 \scriptscriptstyle \pm \scriptstyle .158$ & $0.479 \scriptscriptstyle \pm \scriptstyle .146$ & $0.446 \scriptscriptstyle \pm \scriptstyle .175$ & $0.502 \scriptscriptstyle \pm \scriptstyle .201$   \\ 
\midrule

\multirow{2}{*}{\begin{tabular}[c]{@{}c@{}}DCR \\ (Nearest distance) \end{tabular}} 
& $D_t, S$
& $0.102 \scriptscriptstyle \pm \scriptstyle .078$ &  $0.110 \scriptscriptstyle \pm \scriptstyle .084$ & $0.124 \scriptscriptstyle \pm \scriptstyle .109$ & $0.105 \scriptscriptstyle \pm \scriptstyle .883$ & $0.081 \scriptscriptstyle \pm \scriptstyle .077$ & $0.082 \scriptscriptstyle \pm \scriptstyle .069$  \\ \cmidrule(lr){2-8} 
& $D_r, S$ 
& $0.117 \scriptscriptstyle \pm \scriptstyle .096$ &  $0.104 \scriptscriptstyle \pm \scriptstyle .083$ & $0.132 \scriptscriptstyle \pm \scriptstyle .105$ & $0.129 \scriptscriptstyle \pm \scriptstyle .094$ & $0.102 \scriptscriptstyle \pm \scriptstyle .080$ & $0.094 \scriptscriptstyle \pm \scriptstyle .067$   \\
\midrule

\multirow{2}{*}{\begin{tabular}[c]{@{}c@{}}NNDR \\ (5th percentile distance) \end{tabular}} 
& $D_t, S$ 
& $0.753 \scriptscriptstyle \pm \scriptstyle .226$ & $0.737 \scriptscriptstyle \pm \scriptstyle .204$ &  $0.740 \scriptscriptstyle \pm \scriptstyle .218$ & $0.733 \scriptscriptstyle \pm \scriptstyle .135$ & $0.834 \scriptscriptstyle \pm \scriptstyle .129$ & $0.835 \scriptscriptstyle \pm \scriptstyle .105$  \\ \cmidrule(lr){2-8} 
& $D_r, S$ 
& $0.750 \scriptscriptstyle \pm \scriptstyle .223$ & $0.703 \scriptscriptstyle \pm \scriptstyle .205$ & $0.714 \scriptscriptstyle \pm \scriptstyle .187$ & $0.802 \scriptscriptstyle \pm \scriptstyle .103$ & $0.881 \scriptscriptstyle \pm \scriptstyle .101$ & $0.795 \scriptscriptstyle \pm \scriptstyle .117$    \\
\midrule
\multirow{2}{*}{\begin{tabular}[c]{@{}c@{}}NNDR \\ (Nearest distance) \end{tabular}} 
& $D_t, S$ 
& $0.532 \scriptscriptstyle \pm \scriptstyle .274$ & $0.508 \scriptscriptstyle \pm \scriptstyle .315$ &  $0.496 \scriptscriptstyle \pm \scriptstyle .229$ & $0.517 \scriptscriptstyle \pm \scriptstyle .284$ & $0.542 \scriptscriptstyle \pm \scriptstyle .247$ & $0.522 \scriptscriptstyle \pm \scriptstyle .203$  \\ \cmidrule(lr){2-8} 
& $D_r, S$ 
& $0.530 \scriptscriptstyle \pm \scriptstyle .298$ & $0.498 \scriptscriptstyle \pm \scriptstyle .304$ &  $0.504 \scriptscriptstyle \pm \scriptstyle .209$ & $0.539 \scriptscriptstyle \pm \scriptstyle .263$ & $0.547 \scriptscriptstyle \pm \scriptstyle .229$ & $0.512 \scriptscriptstyle \pm \scriptstyle .255$    \\
\midrule

Groundhog (TPR@1\%FPR) 
& $D_t, \mathsf{A}$ 
& $0.010 \scriptscriptstyle \pm \scriptstyle .002$ & $0.011 \scriptscriptstyle \pm \scriptstyle .001$ &  $0.010 \scriptscriptstyle \pm \scriptstyle .003$ & $0.010 \scriptscriptstyle \pm \scriptstyle .002$ & $0.015 \scriptscriptstyle \pm \scriptstyle .003$ & $0.013 \scriptscriptstyle \pm \scriptstyle .002$  \\
\midrule

TAPAS (TPR@1\%FPR) 
& $D_t, \mathsf{A}$ 
& $0.012 \scriptscriptstyle \pm \scriptstyle .001$ & $0.013 \scriptscriptstyle \pm \scriptstyle .001$ &  $0.011 \scriptscriptstyle \pm \scriptstyle .002$ & $0.009 \scriptscriptstyle \pm \scriptstyle .001$ & $0.030 \scriptscriptstyle \pm \scriptstyle .002$ & $0.020 \scriptscriptstyle \pm \scriptstyle .001$  \\
\midrule

MODIAS (TPR@1\%FPR) 
& $D_t, \mathsf{A}$ 
& $0.011 \scriptscriptstyle \pm \scriptstyle .001$ & $0.011 \scriptscriptstyle \pm \scriptstyle .001$ &  $0.010 \scriptscriptstyle \pm \scriptstyle .002$ & $0.008 \scriptscriptstyle \pm \scriptstyle .001$ & $0.035 \scriptscriptstyle \pm \scriptstyle .002$ & $0.022 \scriptscriptstyle \pm \scriptstyle .001$  \\
\midrule

MDS (ours) 
& $D_t, \mathsf{A}$ 
& $0.031 \scriptscriptstyle \pm \scriptstyle .001$ & $0.046 \scriptscriptstyle \pm \scriptstyle .002$ &  $0.192 \scriptscriptstyle \pm \scriptstyle .003$ & $0.131 \scriptscriptstyle \pm \scriptstyle .002$ & $0.204 \scriptscriptstyle \pm \scriptstyle .001$ & $0.199 \scriptscriptstyle \pm \scriptstyle .001$  \\

\bottomrule
\end{tabular}}

%% file: table/tune_effectiveness.tex
\resizebox{0.8\textwidth}{!}{\begin{tabular}{c|cccccc|ccc}
    \toprule
    \multicolumn{1}{c|}{\multirow{2}{*}{Tuning Objective}} & \multicolumn{6}{c|}{\textbf{Fidelity} $\uparrow$} & \multicolumn{3}{c}{\textbf{Utility $\uparrow$}} \\
    \cmidrule(lr){2-7}  \cmidrule(lr){8-10}
    & TVD & KST& Theil & Pearson & Correlation Ratio &  Wasserstein & Query Errors & MLE & MLA  \\
    \midrule
     MLE$_\text{obj}$ & $2.45$ & $1.52$ & $2.26$ & $2.47$ & $2.61$ & $2.18$ & $2.63$ & $10.58$ & $7.34$ \\ 
    \mymethod & $\mathbf{10.15}$ & $\mathbf{14.83}$ & $\mathbf{11.46}$ & $\mathbf{12.47}$ & $\mathbf{13.83}$ & $\mathbf{13.62}$ & $\mathbf{11.95}$ & $\mathbf{13.06}$ & $\mathbf{13.67}$ \\
    \bottomrule
\end{tabular}}


%% file: 9.3_app_fidelity.tex
\section{Existing Fidelity Metrics and Limitations}
\label{appendix:existing_fidelity_metrics}

\mypara{Low-order Statistics}
Marginals are the workhorses of statistical data analysis and well-established statistics for one(two)-way marginals have been used to assess the quality of synthetic data. 

\textit{Distribution Measurements.}
Total Variation Distance (TVD) and Kolmogorov-Smirnov Test (KST) are used to measure the univariate distribution similarity for categorical and numerical attributes, respectively. The main problem with this approach is the lack of versatility. Each type of marginal requires a distinct statistical measure, which complicates the ability to perform a comprehensive comparison across various attribute types. 

\textit{Correlation Statistics.}
Some researchers use correlation difference, \ie the difference of correlation scores on synthetic and real data, to measure the pairwise distribution similarity. Popular correlation statistics like Theil's uncertainty coefficient~\citep{acml21ctgabgan}, Pearson correlation~\citep{iclr24tabsyn}, and the correlation ratio~\citep{icml23tabddpm} are applied for different types of two-way marginals (categorical, continuous, and mixed).  In addition to the lack of universality, this approach also suffers from the problem that correlation scores capture only limited information about the data distribution. Two attributes may have the same correlation score both in the real data and in the synthetic data, yet their underlying distributions diverge significantly---a phenomenon known as the scale invariance of correlation statistics.

\mypara{Likelihood Fitness} 
\citet{nips19ctgan} assume the input data are generated from some known probabilistic models (\eg Bayesian networks), thus the likelihood of synthetic data can be derived by fitting them to the priors. While likelihood fitness can naturally reflect the closeness of synthetic data to the assumed prior distribution, it is only feasible for data whose priors are known, which is inaccessible for most real-world complex datasets.

\mypara{Evaluator-dependent Metrics}
Probabilistic mean squared error (pMSE)~\citep{snoke2018general} employs a logistic regression discriminator to distinguish between synthetic and real data, using relative prediction confidence as fidelity. 
The effectiveness of pMSE highly relies on the choice of auxiliary discriminator, which requires careful calibration to ensure meaningful comparisons across different datasets and synthesizers.
\citet{icml22faithful} propose $\alpha$-Precision and $\beta$-Recall to quantify how faithful the synthetic data is. 
However, \citet{iclr24tabsyn} finds that $\alpha$-Precision and $\beta$-Recall exhibit a predominantly negative correlation, and it's unclear which one should be used.

\begin{figure}[t]
    \centering
    \vspace{-3mm}
    \includegraphics[width=0.3\textwidth]{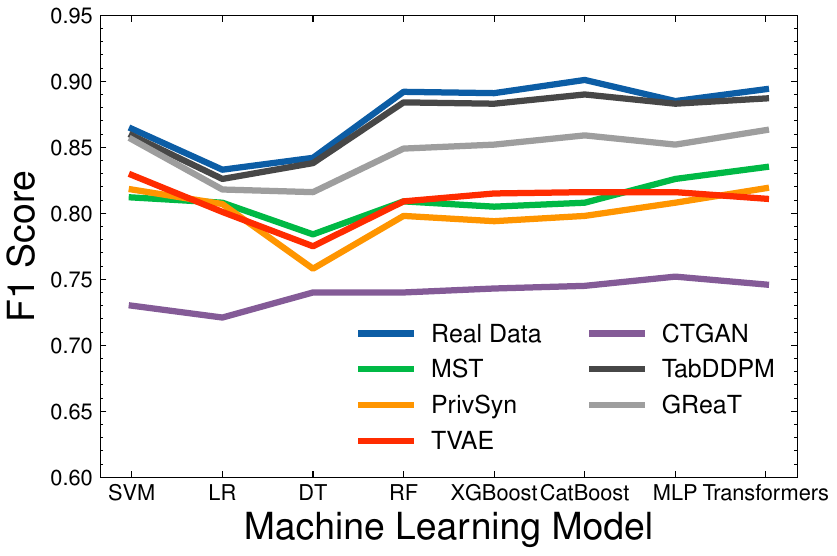}
    \vspace{-3mm}
    \caption{Performance behaviors of HP synthesizers on Magic dataset. ``LR'' denotes Linear Regression, ``DT'' is Decision Tree, and ``RF'' means Random Forest.}
    \vspace{-3mm}
    \label{fig:mle_mla}
\end{figure}



    

%% file: 9.3_app_utlity.tex
\section{Limitations of Machine Learning Efficacy}
\label{appendix:existing_utility_metrics}


We compare the performance of different ML models on the Adult dataset, as illustrated in Figure~\ref{fig:mle_mla}. 
The performance of various synthesizers fluctuates significantly across different ML models, and such variations underscore the impact of the choice of evaluation models. 
For instance, while PrivSyn is ranked third when evaluated using linear regression, it falls to fifth when assessed with decision trees. 
Moreover, directly averaging the performance across all models also fails to capture nuanced performance differences. 
For instance, the mean performance of PrivSyn and TVAE appears nearly identical (0.8 vs. 0.802), whereas MLA more effectively differentiates their relative performance degradation (0.085 vs. 0.075), providing a more reliable assessment.